\documentclass[preprint,12pt]{elsarticle}

\usepackage{amssymb}
\usepackage{amsmath}
\journal{Physica C}

\begin{document}

\begin{frontmatter}



\title{Dependence of the energy and orbital structure of local states in CuO monolayer on Coulomb parameters} %% Article title

%\author{} %% Author name
\author[1]{I.A.~Makarov}
\author[1]{M.M.~Korshunov}
\author[1]{S.G.~Ovchinnikov}
\affiliation[1]{organization={Kirensky Institute of Physics, Federal Research Center KSC SB RAS},%Department and Organization
            addressline={Akademgorodok, 50, bld. 38}, 
            city={Krasnoyarsk},
            postcode={660036},
            country={Russia}}
            
%% Abstract
\begin{abstract}
The dependence of the energies and orbital structure of local states in the CuO monolayer on intra- and interatomic Coulomb interactions on copper and oxygen orbitals is studied. The electronic system is described within the eight-band $p-d$ model in the hole representation with the on-site energies and hopping integrals obtained using density functional theory. CuO cluster multiparticle eigenstates are calculated using exact diagonalization. The difference between the energy dependencies on the Coulomb parameters for the states with the predominant probability density on the $d$-orbital and the states in which hole occupies $p$-orbitals leads to crossover of $d$- and $p$-states. The ground single-hole and two-hole states which determine the electronic structure of the low-energy excitations have the character of $d$- or $p$-orbitals in the different regions of the Coulomb parameters space. The gap between the energies of the dispersionless quasiparticles forming the top of the valence band and conductivity band also have different values in these two regions. The magnitude of this gap and the orbital character of the local multiparticle states change sharply even with an insignificant change in the Coulomb interactions within the boundary region of parameters between the regions in which the local states are formed by the $d$- or $p$-orbitals.
\end{abstract}


%\item There are regions of Coulomb parameters in which the ground local multiparticle states have the character of either purely copper $d$-orbitals or purely oxygen $p$-orbitals and the band gap magnitude varies strongly
%\item The orbital character of the ground local states and the band gap magnitude change significantly when the Coulomb parameters are varied within a narrow range of values between the regions of $d$- and $p$-states

%% Keywords

\end{frontmatter}


\section{Introduction}
\label{Introduction}
The electronic structure of transition metal compounds differ significantly from each other and often exhibit behavior that cannot be described within the single-electron approach. Jan Zaanen and his colleagues have made a prominent contribution to understanding the nature of the electronic states in these compounds~\cite{Zaanen1985,Zaanen1986,Zaanen1990,AnisimovZaanen1991,Liechtenstein1995}. They combined the Mott-Hubbard theory with the concept of charge transfer between transition metal atoms and ligands in order to classify the types of band gaps and electronic states in transition metal oxides, chlorides, sulfides, bromides, iodides, and fluorides. Some of the most interesting and important transition metal compounds in terms of their properties are the high-temperature superconducting (HTSC) cuprates. Significant progress in understanding the electronic states of HTSC cuprates is due to the studies of J. Zaanen and co-workers devoted to the spin and charge ordering~\cite{Zaanen1989,Littlewood1993,Zaanen1998,Zaanen2000,Zaanen2001}. Relatively recently, one of the transition metal oxides, copper oxide, was synthesized as a monolayer in the pores of graphene layers and in a free-standing state~\cite{Kano2017,Yin2017,Kvashnin2019}. The CuO monolayer can potentially have some of the properties of HTSC cuprates since it has much in common with them, and also have its own unique characteristics. However, the CuO monolayer is poorly studied until now. Therefore, it remains unclear whether these compounds exhibit superconductivity, whether they are metals or insulators, and what is the nature of the local states.

The character of the local states and electronic structure of HTSC cuprates have been studied in detail. In undoped cuprates, all 2p oxygen states are fully occupied, and there is one hole on the each copper ${{\rm{d}}_{{x^2} - {y^2}}}$-orbital. The conductivity band are separated from the copper-oxygen states of the valence band by the charge-transfer gap with a size of $2$ eV. Doped holes occupy predominantly oxygen orbitals forming local two-particle Zhang-Rice singlets. It is assumed that these singlets are involved in the process of superconducting pairing due to superexchange interaction. Therefore, any theory describing electronic processes in HTSC cuprates should include single-particle ${{\rm{d}}_{{x^2} - {y^2}}}$ copper and ${\rm{p}}_x$, ${\rm{p}}_y$ oxygen states, as well as the two-particle singlet Zhang-Rice.
 
The crystal structure of the CuO monolayer has common features with CuO$_2$ planes, the most important structural element of HTSC cuprates, such as atomic composition, square crystal lattice, presence of Cu-O-Cu elements in their structure, fullfilled 2p oxygen shells and one hole in the unit cell of the undoped compound. However, the crystal structure and local environment of copper and oxygen atoms differ from cuprates. The oxygen atoms are located between next nearest neighbors of copper atoms, not nearest ones as in HTSC cuprates. The overlap between the orbitals of the nearest neighbors of copper is direct, unlike the CuO$_2$ planes. This feature can lead to a complete change in the energies and orbital structure of local electronic states which, in turn, determines the electronic structure of low-energy excitations and the possibility of realizing a superconducting state.

 This work is devoted to the calculation and investigation of the energies and orbital structure of the local multiparticle states of the CuO monolayer. The study of local states is not only a actual task in itself but also is the basis for obtaining the electronic structure of excitations in the CuO monolayer. Because of its similarity to HTSC cuprates it can be expected that there are strong electronic correlations and strong covalence effects in CuO monolayer. The generalized tight-binding (GTB) method~\cite{Ovchinnikov89,Gavrichkov00,Korshunov05,OvchinnikovValkov} will be used to exactly take into account strong local interaction when constructing quasiparticles that form the electronic structure. In the GTB method, quasiparticles are constructed as a Fermi-type excitations between multiparticle eigenstates of a single cluster with number of holes differing by one. In this work, we will obtain the CuO cluster eigenstates with number of holes $n_h=0,1,2$ using exact diagonalization within eightband $p-d$ model in the hole representation. The $p-d$ model Hamiltonian include the on-site energies, hopping intergrals and different types of Coulomb interactions: intraatomic intraorbital $U_d$, interorbital $V_d$, Hund exchange on copper $J_d$, intraatomic interaction on oxygen $U_p$ and interatomic copper-oxygen $V_{pd}$. The values of on-site energies and hopping intergrals are obtained using first-principles calculations within the framework of the local density functional method~\cite{Slobodchikov2023}. The values of the Coulomb parameters are assumed to be close to their values in cuprates calculated within the density functional theory (DFT) approach in the J. Zaanen's work~\cite{Gunnarsson1989} and other calculations using the local-density-functional and constrained density-functional approaches~\cite{Schluter88,Schluter89,Hybertsen89,Hybertsen90,Mahan90,Grant1992,Anisimov1991,Anisimov1992}. The calculated Coulomb parameters lie in a wide range of values. Due to the lack of information regarding the characteristics of the electronic structure, there are no guidelines for estimating the realistic magnitudes of Coulomb interactions. It is unclear how sensitive the structure and energies of local states are to the Coulomb parameters. It is likely that small changes in the Coulomb parameters can lead to significant reconstruction of the top of the valence band, a region key to understanding low-energy properties and possible superconductivity. Therefore, it is necessary to investigate the electronic structure of local states in the CuO monolayer in a wide range of Coulomb parameters. 
 
The paper consists of five sections. Section 2 presents the Hamiltonian of the eight-band $p-d$ model in the electron and hole representations. Section 3 is devoted to the structure and orbital composition of the zero-, single- and two-hole CuO cluster eigenstates. Section 4 contains the dependencies of energies of CuO cluster eigenstates and type of orbitals with maximal probability density in the ground states on Coulomb parameters. Section 5 describes the dependence of the gap magnitude between the energies of quasiparticle excitations forming conductivity and valence bands on the Coulomb parameters. The conclusion contains the main results of the paper.   
 
\section{Hamiltonian of a single cluster within the eightband $p-d$ model}
\label{sec:hamiltonian}

\begin{figure}
\centering
\includegraphics[width=0.3\linewidth]{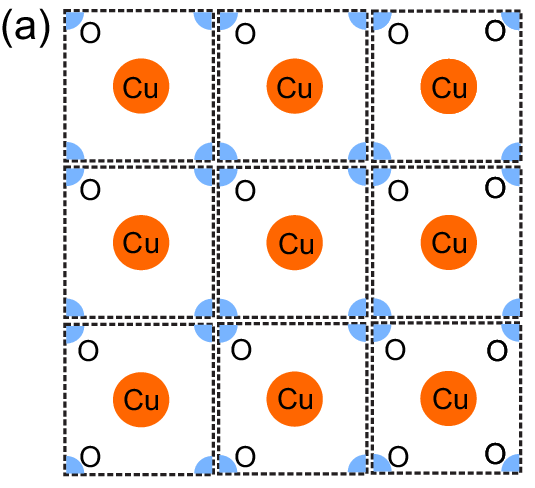}
\includegraphics[width=0.69\linewidth]{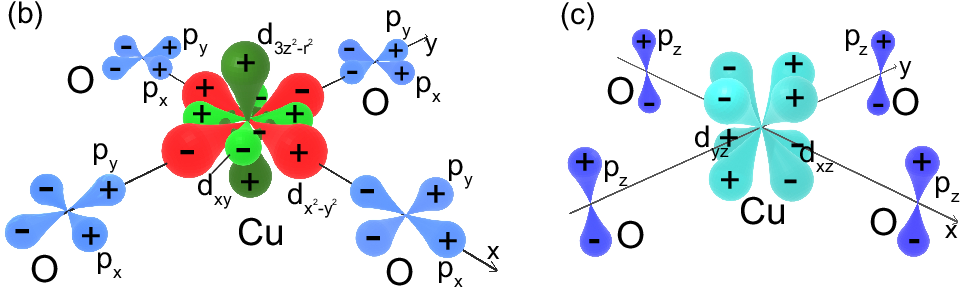}
\caption{(a) Division of the CuO monolayer into CuO clusters. (b), (c) Copper and oxygen orbitals of the CuO cluster. Pluses and minuses signs are signs of the phases of the wave function.}\label{fig:cluster}
\end{figure}
In the GTB method, the crystal lattice of CuO monolayer is divided into separate clusters (Fig.1). We choose unit cell CuO consisting of Cu atom surrounded four O atoms as a single cluster. Basis electron orbitals includes five $3d$ orbitals of copper ${{\rm{d}}_{{x^2} - {y^2}}}$, ${{\rm{d}}_{3{z^2} - {r^2}}}$, ${{\rm{d}}_{xy}}$, ${{\rm{d}}_{xz}}$, ${{\rm{d}}_{yz}}$ and three $2p$ orbitals ${{\rm{p}}_x}$, ${{\rm{p}}_y}$, ${{\rm{p}}_z}$ on each oxygen atom. To describe the electronic system, the eightband $p-d$ model will be used
\begin{eqnarray}
H &=& \sum\limits_{\mathbf{r} \lambda \sigma } {\left( {{\varepsilon _{\mathbf{r}\lambda} } - \mu } \right){n_{\mathbf{r}\lambda \sigma }}}  + \sum\limits_{\mathbf{r} \mathbf{r}' \lambda \lambda'} {{t_{\lambda \lambda '}}\left( {{{\bf{R}}_{\lambda \lambda '}}} \right)c_{\mathbf{r}\lambda \sigma }^\dag {c_{\mathbf{r}' \lambda '\sigma }}} + {H_U}, \nonumber\\
{H_U} &=& \sum\limits_{\mathbf{r} \lambda}  {{U_\lambda }{n_{\mathbf{r}\lambda  \downarrow }}{n_{\mathbf{r}\lambda  \uparrow }}}  + \frac{1}{2}\sum\limits_{\mathbf{r} \lambda  \ne \lambda' \sigma \sigma'} {{V_{\lambda \lambda '}}{n_{\mathbf{r}\lambda \sigma }}{n_{\mathbf{r}\lambda '\sigma '}}} + \nonumber\\
&+& \frac{1}{2}\sum\limits_{\mathbf{r} \lambda \ne \lambda' \sigma \sigma'} {{J_{\lambda \lambda '}}c_{\mathbf{r}\lambda \sigma }^\dag {c_{\mathbf{r}\lambda \sigma '}}c_{\mathbf{r}\lambda '\sigma '}^\dag {c_{\mathbf{r}\lambda '\sigma }}}.
\label{eq:Ham_el}
\end{eqnarray}

Here $\lambda$, $\lambda '$ are the indices of atomic copper $d$-orbitals and oxygen $p$-orbitals in the CuO monolayer, ${c_{\mathbf{r}\lambda \sigma }}$ is the annihilation operator for the electron with the spin projection $\sigma$ on the orbital $\lambda$ of the site $\mathbf{r}$, ${c_{\mathbf{r}\lambda \sigma }}={{\rm{d}}_{\mathbf{r}\lambda \sigma}}$ or ${{\rm{p}}_{\mathbf{r}\lambda \sigma}}$, ${{n_{\mathbf{r}\lambda \sigma }}}={c_{\mathbf{r}\lambda \sigma }^\dag {c_{\mathbf{r}\lambda \sigma }}}$. The parameter ${\varepsilon _{\mathbf{r}\lambda }}$ is the on-site energy of an electron on the orbital ${\lambda }$, ${\varepsilon _{\mathbf{r}\lambda} } = {\varepsilon _{d \lambda}}$ for the $d$-orbitals and ${\varepsilon _{\mathbf{r}\lambda} } = {\varepsilon _{p \lambda}}$ for the $p$-orbitals. ${t_{\lambda \lambda '}}\left( {{{\bf{R}}_{\lambda \lambda '}}} \right)$ is the hopping integrals between orbitals $\lambda$ and $\lambda '$, the vector ${\bf{R}}_{\lambda \lambda '} = \mathbf{r} - \mathbf{r}'$ connects atoms at the sites $\mathbf{r}$ and $\mathbf{r}'$. The Cu-Cu, O-O hoppings are considered between five nearest neighbors and the Cu-O hoppings are between four nearest neighbors (Tables~\ref{table1}-\ref{table4} in~\ref{app:parameters}). The intraatomic intraorbital Coulomb interaction parameter ${U_\lambda }$ take the value ${U_d }$ for copper orbitals and $U_p$ for oxygen orbitals. If the interorbital Coulomb parameter ${V_{\lambda \lambda '}}$ characterizes interaction between charge carriers on the orbitals $\lambda$ and $\lambda '$ of the same atom, then (${V_{\lambda \lambda '}}={V_d }$ for Cu and $V_p=U_p$ for O). The parameter ${V_{\lambda \lambda '}}$ turns into the interatomic Coulomb interaction $V_{pd}$ in the case when one of the orbitals $\lambda$, $\lambda '$ is located on Cu and another is on O. The parameter ${J_{\lambda \lambda '}}$ is the Hund exchange between carriers on the $d$-orbitals of the copper atom (${J_{\lambda \lambda '}}=J_d$, $J_p=0$).

The Hamiltonian of eightband $p-d$ model should be represented as a sum of intracluster interactions and the Hamiltonian of intercluster interactions:
\begin{equation}\label{eq:divided_Ham}
H=\sum\limits_{\bf{f}} {H_{\bf{f}}^{ c }} + \sum\limits_{{\bf{fg}}}{H_{{\bf{fg}}}^{cc}},
\end{equation}
where ${\bf{f}}$ and ${\bf{f}}$ are the coordinates of CuO clusters. To isolate Hamiltonian of intracluster interactions inside CuO cluster we need orthogonalize oxygen states since each oxygen atom belongs to four clusters. The transformation from atomic oxygen orbitals ${{\rm{p}}_x}$, ${{\rm{p}}_y}$, ${{\rm{p}}_z}$ to new molecular oxygen orbitals ${\alpha _{\bf{k}}}$, ${\beta _{\bf{k}}}$, ${\gamma _{\bf{k}}}$ in the k-space was made (\ref{app:str_factors}). It is more convenient to work with multielectron local states when they are written in a hole representation against the background of a state completely filled with electrons ${\left| {FB} \right\rangle _e}$. The electron annihilation (creation) operators ${c_{\lambda \sigma }}$ ($c_{\lambda \sigma }^\dag$) are replaced by the creation (annihilation) operators of holes $h_{\lambda \bar \sigma }^\dag$ ($h_{\lambda \bar \sigma }$):
\begin{equation}
h_{\lambda \bar \sigma }^\dag {\left| 0 \right\rangle _h} = {c_{\lambda \sigma }}{\left| {FB} \right\rangle _e}, \;\;\;
{h_{\lambda \bar \sigma }}{\left| 0 \right\rangle _h} = c_{\lambda \sigma }^\dag {\left| {FB} \right\rangle _e},
\end{equation}
where ${\left| 0 \right\rangle _h}$ is the vacuum state of holes. Further, we will write the operators of creation and annihilation of a hole in italics: ${d_{{{x^2} - {y^2}}\sigma }}$, ${d_{{3{z^2} - {r^2}}\sigma }}$, ${d_{xy\sigma }}$, ${d_{xz\sigma }}$, ${d_{yz\sigma }}$, ${p_{x\sigma }}$, ${p_{y\sigma }}$, ${p_{z\sigma }}$. The Hamiltonian of a single cluster in the hole representation has the form:
\begin{eqnarray}\label{eq:Ham_mo}
&&H^{ c } = {\varepsilon _0} - \\\nonumber
&&\sum\limits_{\zeta \sigma } {\left[ {\left( {\varepsilon _{d\zeta }^{\left( h \right)} - \mu } \right)d_{\zeta \sigma }^\dag {d_{\zeta \sigma } }+\sum\limits_{\zeta '} {{t_{\zeta \zeta '}}d_{\zeta \sigma }^\dag {d_{\zeta '\sigma }} + \sum\limits_{j} {\left( {\kappa _{0}^{\left( {\zeta {\rho _j}} \right)}d_{\zeta \sigma }^\dag {\rho _{j\sigma }} + h.c.} \right)} } } \right]}  - \\\nonumber
&& \sum\limits_{ij\sigma } {\left[ {{\delta _{ij}}\left( {\nu _{0}^{\left( i \right)} - \mu } \right) + \left( {1 - {\delta _{ij}}} \right)\nu _{0}^{\left( {ij} \right)}} \right]\rho _{i\sigma }^\dag {\rho _{j\sigma }}}  +\\\nonumber
&& \sum\limits_{ii'jj'} {{U_p}\Psi _{0000}^{{\rho _i}{\rho _{i'}}{\rho _j}{\rho _{j'}}}\rho _{i\sigma }^\dag {\rho _{i'\sigma }}\rho _{j\sigma '}^\dag {\rho _{j'\sigma '}}}  + \\\nonumber
&&\sum\limits_{\zeta \sigma } {\left\{ {\frac{1}{2}{U_d}n_{\zeta \sigma }^{\left( d \right)}n_{\zeta \sigma }^{\left( d \right)} + \frac{1}{2}\sum\limits_{\zeta '\sigma '} {\left[ {{V_d}n_{\zeta \sigma }^{\left( d \right)}n_{\zeta '\sigma '}^{\left( d \right)} - {J_d}d_{\zeta \sigma }^\dag {d_{\zeta \sigma '}}d_{\zeta '\sigma '}^\dag {d_{\zeta '\sigma }}} \right]} } \right\}} +\\\nonumber
&&\sum\limits_{\zeta \sigma } {\sum\limits_{{\bf{gg'}}ij\sigma '} {{V_{pd}}\Phi _{000}^{{\rho _i}{\rho _j}}d_{\zeta \sigma }^\dag {d_{\zeta \sigma }}\rho _{i\sigma '}^\dag {\rho _{j\sigma '}}} }.
\end{eqnarray}
Here ${d_{\zeta \sigma }}$ and ${\rho _{j\sigma }}$ are the annihilation operators of the hole with spin projection $\sigma$ in the five copper $d$-orbitals $\zeta$ and three molecular oxygen orbitals ${\rho _1}\equiv\alpha$, ${\rho _2}\equiv\beta$, ${\rho _3}\equiv\gamma$, respectively. The general definitions of energy parameters ${\nu _{{\bf{fg}}}^{\left( i \right)}}$, ${\nu _{{\bf{fg}}}^{\left( {ij} \right)}}$, ${\kappa _{{\bf{fg}}}^{\left( {\lambda {\rho _j}} \right)}}$, ${\Phi _{{\bf{fgg'}}}^{{\rho _i}{\rho _j}}}$, $\Psi _{{\bf{ff'gg'}}}^{{\rho _i}{\rho _{i'}}{\rho _j}{\rho _{j'}}} $ are given in~\ref{app:str_factors}. The term ${\varepsilon _0}$ is the energy of the completely filled electron $3d$-shells of copper and $2p$-shells of oxygen (vacuum state of holes) on all atoms of the CuO cluster. The on-site hole energy $\varepsilon _{d\zeta }^{\left( h \right)}$ ($\varepsilon _{p\lambda }^{\left( h \right)}$ ) is the energy by which ${\varepsilon _0}$ decreases when a hole appears (or electron disappears) in the corresponding copper ${d_{\zeta} }$ (${p_{\lambda} }$) orbital:
\begin{eqnarray}\label{eq:edh}
{\varepsilon _{d\zeta }^{\left( h \right)}}= {\varepsilon _{d \zeta}} + {{{\tilde U}_d }},\\\nonumber
\left({\varepsilon _{p\lambda }^{\left( h \right)}}= {\varepsilon _{p \lambda}} + {{{\tilde U}_p }}\right),
\end{eqnarray}
where ${\varepsilon _{d \zeta}}$ (${\varepsilon _{p \lambda}}$) is the on-site electron energy, ${{{\tilde U}_d }} = {U_d} + 4{l_{Cu}}\left( {{V_d} + {J_d}} \right) + 2\left( {2{l_O} + 1} \right){z_{dp}}{V_{pd}}$ (${{\tilde U}_p} = \left( {4{l_{Cu}} + 1} \right){U_p} + 2\left( {2{l_{Cu}} + 1} \right){z_{pd}}{V_{pd}}$) is the sum of the energies of the intraorbital, interorbital, and interatomic interactions in which given electron was involved, $l_{R}$ is the orbital number of the atom $R$, $z_{dp}=4$ is the number of nearest neighbors. The sum of the Coulomb interactions ${{{\tilde U}_d }}$ (${{{\tilde U}_p }}$) in the definition (\ref{eq:edh}) means that when a hole is added to the $d_{\zeta}$ (${p_{\lambda} }$) orbital, the energy of the system decreases not only by the energy of the electron, but also by the magnitude of all interactions in which it participates. The on-site hole energies ${\varepsilon _{p\lambda }^{\left( h \right)}}$ are contained in the parameters ${\nu^{\left( i \right)}}$. Each term ${t_{\zeta \zeta '}}$ is the hopping integral between the $d$-orbitals $\zeta$ and $\zeta '$, $\nu _{0}^{\left( {ij} \right)}$ is the hopping integral between the molecular oxygen orbitals ${\rho _{i\sigma }}$ and ${\rho _{j\sigma }}$, $\kappa _{0}^{\left( {\zeta {\rho _j}} \right)}$ is the hopping integral between the atomic $d$-obital $\zeta$ and the molecular $p$-orbital ${\rho _{j\sigma }}$. The values of the on-site electron energies, hopping integrals are given in~\ref{app:parameters}.
 
\section{CuO cluster multiparticle eigenstates}
\begin{figure}
\centering
\includegraphics[width=0.7\linewidth]{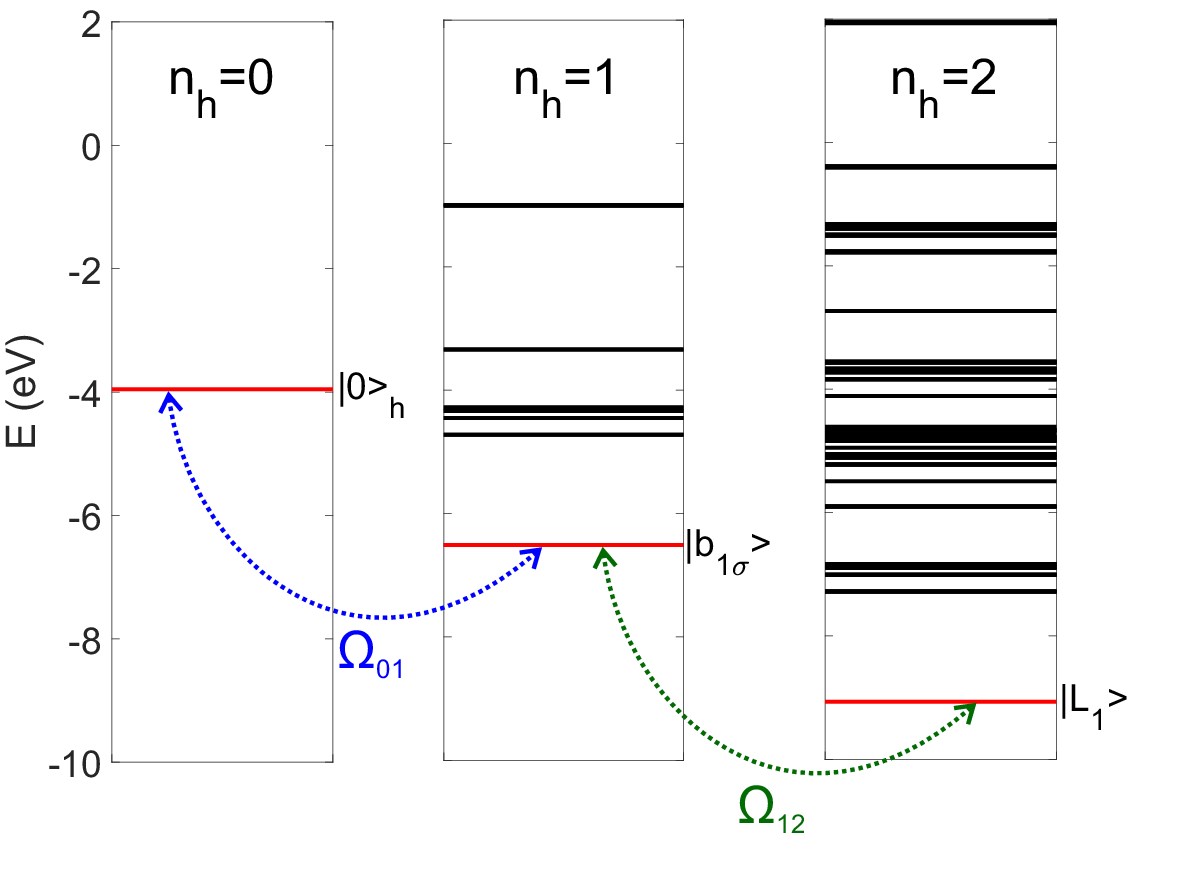}
\caption{The energy levels of zero-, single-, two-hole eigenstates of the CuO cluster without Coulomb interactions}\label{fig:levels}
\end{figure}
%\begin{figure}[t]%% placement specifier
%%% Use \includegraphics command to insert graphic files. Place graphics files in 
%%% working directory.
%\centering%% For centre alignment of image.
%\includegraphics{example-image-a}
%%% Use \caption command for figure caption and label.
%\caption{Figure Caption}\label{fig1}
%%% https://en.wikibooks.org/wiki/LaTeX/Importing_Graphics#Importing_external_graphics
%\end{figure}

\begin{figure}
\centering
\includegraphics[width=0.45\linewidth]{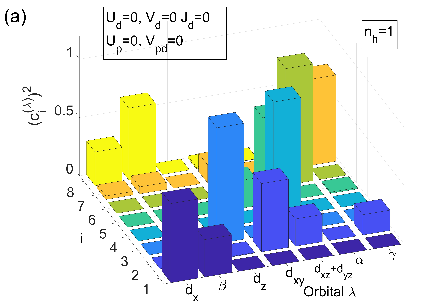}
\includegraphics[width=0.45\linewidth]{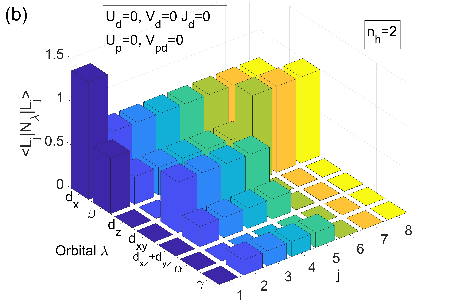}
\caption{(Color online) Distribution of the probability density of hole in eight eigenstates of a CuO cluster with the number of holes (a) ${n_h} = 1$ and (b) ${n_h} = 2$. The hole density in the $\lambda$ orbital of a cluster with one hole is characterized by the square of the probability amplitude ${\left( {c_m^{\left( \lambda \right)}} \right)^2}$ in the wave function $\left| {b_m} \right\rangle$, and, in the case with two holes, it is determined by the average number of particles $\left\langle {{L_j}} \right|{N_\lambda }\left| {{L_m}} \right\rangle $ at a given orbital, where ${N_\lambda } = \sum\limits_\sigma {a_{\lambda \sigma }^\dag {a_{\lambda \sigma }}} $ is the particle number operator.}
\label{fig:eigenstates_str}
\end{figure}

The cluster eigenstates $\left| {m} \right\rangle$ are obtained as a result of solving the stationary Schrodinger equation:
\begin{equation}\label{eq:Schrod}
H^{c}\left| m \right\rangle  = E_m^{\left( n_h \right)}\left| m \right\rangle,
\end{equation}
where $E_m^{\left( n_h \right)}$ are the energies of the cluster eigenstates. The exact diagonalization procedure is performed for a cluster with ${n_h} = 0,1,2$ holes in order to obtain the states into which the cluster transform from the stoichiometric composition with one hole under elementary Fermi excitations that create or annihilate one hole. It is these quasiparticle excitations that will eventually form the electronic structure of the CuO monolayer after taking into account their dispersion. The many-particle eigenstates of the cluster are the one vacuum state with zero holes ${\left| 0 \right\rangle  _h}$, the sixteen single-hole states $\left| {b_{m\sigma }} \right\rangle $ (eight doublets), and the $120$ two-hole states $\left| {{L_m}} \right\rangle $ (Fig.~\ref{fig:cluster}). In the hole representation, the states with the lowest energy will form the highest energy bands since they lie closest to the completely filled shells. Therefore, in order to understand the nature of the states at the top of the valence band and in the conduction band, it is necessary first of all to analyze the structure of the ground multiparticle states of the CuO cluster (Fig.~\ref{fig:levels}, red lines) and transitions between them.

First, we consider the CuO eigenstates in the case without taking into account Coulomb interactions. The single-hole eigenstates are the superposition of copper and oxygen states corresponding to the electronic configurations $d^9p^6$ and $d^{10}p^5$. The energy of a single-hole state consists of the on-site energies of the orbitals that compose this state, hopping integrals between these orbitals and intraatomic and interatomic Coulomb interaction. Electron on-site energies of copper $d$-orbitals is larger than that of oxygen $p$-orbitals. In the hole representation, on the contrary, the states corresponding electronic configuration $d^9p^6$ have lower energies than the hole states corresponding the electronic configuration $d^{10}p^5$. The hole states with the predominant probability density on $d$-orbitals (the states with $m=1-5$ in Fig.~\ref{fig:eigenstates_str}a)
\begin{eqnarray}\label{eq:nh1_d}
&&\left| {{b_{1\sigma }}} \right\rangle  = c_1^{\left( {d_{{x^2} - {y^2}}} \right)}\left| {{d_{{{x^2} - {y^2}}\sigma }}} \right\rangle  + c_1^{\left( \beta  \right)}\left| {{\beta _\sigma }} \right\rangle ,\,\,\sigma  =  \downarrow , \uparrow ,\\
&&\left| {{b_{2\sigma }}} \right\rangle  = c_2^{\left( {{d_{xy}}} \right)}\left| {{d_{xy\sigma }}} \right\rangle  + c_2^{\left( {{d_{xz}}} \right)}\left| {{d_{xz\sigma }}} \right\rangle  + c_2^{\left( {{d_{yz}}} \right)}\left| {{d_{yz\sigma }}} \right\rangle  + c_2^{\left( \gamma  \right)}\left| {{\gamma _\sigma }} \right\rangle \\
&&\left| {{b_{3\sigma }}} \right\rangle  = c_2^{\left( {{d_{3{z^2} - {r^2}}}} \right)}\left| {{d_{3{z^2} - {r^2}\sigma }}} \right\rangle \\
&&\left| {{b_{4\sigma }}} \right\rangle  = c_4^{\left( {{d_{xz}}} \right)}\left| {{d_{xz\sigma }}} \right\rangle  + c_4^{\left( {{d_{yz}}} \right)}\left| {{d_{yz\sigma }}} \right\rangle \\
&&\left| {{b_{5\sigma }}} \right\rangle  = c_5^{\left( {{d_{xy}}} \right)}\left| {{d_{xy\sigma }}} \right\rangle  + c_5^{\left( {{d_{xz}}} \right)}\left| {{d_{xz\sigma }}} \right\rangle  + c_5^{\left( {{d_{yz}}} \right)}\left| {{d_{yz\sigma }}} \right\rangle,
\end{eqnarray}
where the hole states are defined as $\left| {{d_{{{x^2} - {y^2}} \sigma }}} \right\rangle  = d_{{{x^2} - {y^2}} \sigma }^\dag {\left| 0 \right\rangle _h}$, $\left| {{\beta _\sigma }} \right\rangle  = \beta _\sigma ^\dag {\left| 0 \right\rangle _h}$, form a separate group of low-energy states characterized by similar behavior with changing Coulomb parameters, as will be seen in the next section. The electronic ${{\rm{d}}_{{x^2} - {y^2}}}$-orbital has the highest energy among $d$-orbitals and it also have the largest overlap with the oxygen $p_x$-, $p_y$-orbitals. Therefore, the ground hole state of the cluster in the single-hole Hilbert space sector is the doublet of the same orbitals as in cuprates. The probability to find a hole in the ground state $\left| {{b_1}} \right\rangle $ on the copper ${d_{{x^2} - {y^2}}}$-orbital in the case without Coulomb interactions is higher than on the oxygen $\beta$-orbital (Fig.~\ref{fig:eigenstates_str}a). The first excited single-hole state which is characterized by the predominant hole density in the copper ${d_{xy}}$-orbital and the smaller one in the copper ${d_{xz}}$-, ${d_{yz}}$- and oxygen $\gamma$-orbitals (Fig.~\ref{fig:eigenstates_str}a, $m = 2$) is located $1.783$~eV above the ground state. In the second excited state, the hole is almost completely localized in the ${d_{3{z^2} - {r^2}}}$-orbital, and in the third excited state it is located in the ${d_{xz}}$-, ${d_{yz}}$-orbitals (Fig.~\ref{fig:eigenstates_str}a).
 
 The another group of states consists of the states with the predominant probability density on the molecular oxygen orbitals $\alpha$, $\beta$, $\gamma$ (the states with $m=6-8$ in Fig.~\ref{fig:eigenstates_str}):
\begin{eqnarray}\label{eq:nh1_p}
&&\left| {{b_{6\sigma }}} \right\rangle  = c_6^{\left( \alpha  \right)}\left| {{\alpha _\sigma }} \right\rangle \\
&&\left| {{b_{7\sigma }}} \right\rangle  = c_7^{\left( \gamma  \right)}\left| {{\gamma _\sigma }} \right\rangle  + c_7^{\left( {{d_{xy}}} \right)}\left| {{d_{xy\sigma }}} \right\rangle  + c_7^{\left( \beta  \right)}\left| {{\beta _\sigma }} \right\rangle  + c_7^{\left( {{d_{{x^2} - {y^2}}}} \right)}\left| {{d_{{x^2} - {y^2}\sigma }}} \right\rangle  + \\
&&c_7^{\left( {{d_{xz}}} \right)}\left| {{d_{xz\sigma }}} \right\rangle  + c_7^{\left( {{d_{yz}}} \right)}\left| {{d_{yz\sigma }}} \right\rangle  \\\nonumber
&&\left| {{b_{8\sigma }}} \right\rangle  = c_8^{\left( \beta  \right)}\left| {{\beta _\sigma }} \right\rangle  + c_8^{\left( {{d_{{x^2} - {y^2}}}} \right)}\left| {{d_{{x^2} - {y^2}\sigma }}} \right\rangle  + c_8^{\left( \gamma  \right)}\left| {{\gamma _\sigma }} \right\rangle .
\end{eqnarray}
There are both pure oxygen ($\left| {{b_6}} \right\rangle $) and the states $b_7$, $b_8$ with a small admixture of $d$-orbitals which are antibonding components of $\left| {{b_2}} \right\rangle $, $\left| {{b_1}} \right\rangle $, respectively. All single-hole states can be divided into the two groups according to the distribution of the predominant hole density over $d$- and $p$-orbitals. If the hole is mainly on the copper $d$-orbitals, then this single-hole state relates to the first group (the group of $d$-states). The second group (the group of $p$-states) consists of the single-hole states with the predominant probability hole density on the oxygen $p$-orbitals. 

The two-hole eigenstates are the superposition of $dd$, $pp$ and $dp$ states corresponding to electronic configurations $d^8p^6$, $d^{10}p^4$ and $d^9p^5$. The ground two-hole state includes the singlet $\left| {{d_{{{x^2} - {y^2}} \downarrow }}{\beta _ \uparrow }} \right\rangle - \left| {{d_{{{x^2} - {y^2}} \uparrow }}{\beta _ \downarrow }} \right\rangle $ which is an analogue of the Zhang-Rice singlet in HTSC cuprates, the state with two holes on the copper orbital and the state with two holes on the oxygen orbitals:
\begin{eqnarray}\label{eq:ZRS}
&&\left| {L_1^S} \right\rangle  = L_1^{\left( {d_{{x^2} - {y^2}}\beta } \right)}\left( {\left| {{d_{{{x^2} - {y^2}} \downarrow }}{\beta _ \uparrow }} \right\rangle  - \left| {{d_{{{x^2} - {y^2}} \uparrow }}{\beta _ \downarrow }} \right\rangle } \right) +\\\nonumber
&& L_1^{\left( {d_{{x^2} - {y^2}}d_{{x^2} - {y^2}}} \right)}\left| {{d_{{{x^2} - {y^2}} \downarrow }}{d_{{{x^2} - {y^2}} \uparrow }}} \right\rangle  + L_1^{\left( {\beta \beta } \right)}\left| {{\beta _ \downarrow }{\beta _ \uparrow }} \right\rangle 
\end{eqnarray}
The first excited state is fourfold degenerate in energy and includes the singlet
\begin{eqnarray}\label{eq:singlet_1exc}
&&\left| {L_2^S} \right\rangle  = L_2^{\left( {d_{{x^2} - {y^2}}d_{xy}} \right)}\left( {\left| {{d_{{{x^2} - {y^2}} \downarrow }}{d_{xy \uparrow }}} \right\rangle  - \left| {{d_{{{x^2} - {y^2}} \uparrow }}{d_{xy \downarrow }}} \right\rangle } \right) +\\\nonumber
&& L_2^{\left( {d_{{x^2} - {y^2}}d_{xz}} \right)}\left( {\left| {{d_{{{x^2} - {y^2}} \downarrow }}{d_{xz \uparrow }}} \right\rangle  - \left| {{d_{{{x^2} - {y^2}} \uparrow }}{d_{xz \downarrow }}} \right\rangle } \right) + \\\nonumber
&&L_2^{\left( {d_{{x^2} - {y^2}}d_{yz}} \right)}\left( {\left| {{d_{{{x^2} - {y^2}} \downarrow }}{d_{yz \uparrow }}} \right\rangle  - \left| {{d_{{{x^2} - {y^2}} \uparrow }}{d_{yz \downarrow }}} \right\rangle } \right) +\\\nonumber
&& L_2^{\left( {d_{{x^2} - {y^2}}\gamma } \right)}\left( {\left| {{d_{{{x^2} - {y^2}} \downarrow }}{\gamma _ \uparrow }} \right\rangle  - \left| {{d_{{{x^2} - {y^2}} \uparrow }}{\gamma _ \downarrow }} \right\rangle } \right) + \\\nonumber
&&L_2^{\left( {\beta d_{xy}} \right)}\left( {\left| {{\beta _ \downarrow }{d_{xy \uparrow }}} \right\rangle  - \left| {{\beta _ \uparrow }{d_{xy \downarrow }}} \right\rangle } \right) + L_2^{\left( {\beta d_{xz}} \right)}\left( {\left| {{\beta _ \downarrow }{d_{xz \uparrow }}} \right\rangle  - \left| {{\beta _ \uparrow }{d_{xz \downarrow }}} \right\rangle } \right) + \\\nonumber
&&L_2^{\left( {\beta d_{yz}} \right)}\left( {\left| {{\beta _ \downarrow }{d_{yz \uparrow }}} \right\rangle  - \left| {{\beta _ \uparrow }{d_{yz \downarrow }}} \right\rangle } \right) + L_2^{\left( {\beta \gamma } \right)}\left( {\left| {{\beta _ \downarrow }{\gamma _ \uparrow }} \right\rangle  - \left| {{\beta _ \uparrow }{\gamma _ \downarrow }} \right\rangle } \right)
\end{eqnarray}
and the triplet components
\begin{eqnarray}\label{eq:tripletT0_1exc}
&&\left| {L_3^{T0}} \right\rangle  = L_3^{\left( {d_{{x^2} - {y^2}}d_{xy}} \right)}\left( {\left| {{d_{{{x^2} - {y^2}} \downarrow }}{d_{xy \uparrow }}} \right\rangle  + \left| {{d_{{{x^2} - {y^2}} \uparrow }}{d_{xy \downarrow }}} \right\rangle } \right) +\\\nonumber
&& L_3^{\left( {d_{{x^2} - {y^2}}d_{xz}} \right)}\left( {\left| {{d_{{{x^2} - {y^2}} \downarrow }}{d_{xz \uparrow }}} \right\rangle  + \left| {{d_{{{x^2} - {y^2}} \uparrow }}{d_{xz \downarrow }}} \right\rangle } \right) + \\\nonumber
&&L_3^{\left( {d_{{x^2} - {y^2}}d_{yz}} \right)}\left( {\left| {{d_{{{x^2} - {y^2}} \downarrow }}{d_{yz \uparrow }}} \right\rangle  + \left| {{d_{{{x^2} - {y^2}} \uparrow }}{d_{yz \downarrow }}} \right\rangle } \right) +\\\nonumber
&&L_3^{\left( {d_{{x^2} - {y^2}}\gamma } \right)}\left( {\left| {{d_{{{x^2} - {y^2}} \downarrow }}{\gamma _ \uparrow }} \right\rangle  + \left| {{d_{{{x^2} - {y^2}} \uparrow }}{\gamma _ \downarrow }} \right\rangle } \right) + \\\nonumber
&&L_3^{\left( {\beta d_{xy}} \right)}\left( {\left| {{\beta _ \downarrow }{d_{xy \uparrow }}} \right\rangle  + \left| {{\beta _ \uparrow }{d_{xy \downarrow }}} \right\rangle } \right) + L_3^{\left( {\beta d_{xz}} \right)}\left( {\left| {{\beta _ \downarrow }{d_{xz \uparrow }}} \right\rangle  + \left| {{\beta _ \uparrow }{d_{xz \downarrow }}} \right\rangle } \right) + \\\nonumber
&&L_3^{\left( {\beta d_{yz}} \right)}\left( {\left| {{\beta _ \downarrow }{d_{yz \uparrow }}} \right\rangle  + \left| {{\beta _ \uparrow }{d_{yz \downarrow }}} \right\rangle } \right) + L_3^{\left( {\beta \gamma } \right)}\left( {\left| {{\beta _ \downarrow }{\gamma _ \uparrow }} \right\rangle  + \left| {{\beta _ \uparrow }{\gamma _ \downarrow }} \right\rangle } \right)
\end{eqnarray}
\begin{eqnarray}\label{eq:tripletT_1_1exc}
&&\left| {L_4^{T - 1}} \right\rangle  = L_4^{\left( {d_{{x^2} - {y^2}}d_{xy}} \right)}\left| {{d_{{{x^2} - {y^2}} \downarrow }}{d_{xy \downarrow }}} \right\rangle  +\\\nonumber
&& L_4^{\left( {d_{{x^2} - {y^2}}d_{xz}} \right)}\left| {{d_{{{x^2} - {y^2}} \downarrow }}{d_{xz \downarrow }}} \right\rangle  + L_4^{\left( {d_{{x^2} - {y^2}}d_{yz}} \right)}\left| {{d_{{{x^2} - {y^2}} \downarrow }}{d_{yz \downarrow }}} \right\rangle  +\\\nonumber
&& L_4^{\left( {d_{{x^2} - {y^2}}\gamma } \right)}\left| {{d_{{{x^2} - {y^2}} \downarrow }}{{\rm{\gamma }}_ \downarrow }} \right\rangle  +\\\nonumber 
&&L_4^{\left( {\beta d_{xy}} \right)}\left| {{\beta _ \downarrow }{d_{xy \downarrow }}} \right\rangle  + L_4^{\left( {\beta d_{xz}} \right)}\left| {{\beta _ \downarrow }{d_{xz \downarrow }}} \right\rangle  + L_4^{\left( {\beta dyz} \right)}\left| {{\beta _ \downarrow }{d_{yz \downarrow }}} \right\rangle  + L_4^{\left( {\beta \gamma } \right)}\left| {{\beta _ \downarrow }{\gamma _ \downarrow }} \right\rangle  
\end{eqnarray}
\begin{eqnarray}\label{eq:tripletT1_1exc}
&&\left| {L_5^{T1}} \right\rangle  = L_5^{\left( {d_{{x^2} - {y^2}}d_{xy}} \right)}\left| {{d_{{{x^2} - {y^2}} \uparrow }}{d_{xy \uparrow }}} \right\rangle  + L_5^{\left( {d_{{x^2} - {y^2}}d_{xz}} \right)}\left| {{d_{{{x^2} - {y^2}} \uparrow }}{d_{xz \uparrow }}} \right\rangle  +\\\nonumber
&& L_5^{\left( {d_{{x^2} - {y^2}}d_{yz}} \right)}\left| {{d_{{{x^2} - {y^2}} \uparrow }}{d_{yz \uparrow }}} \right\rangle  + L_5^{\left( {d_{{x^2} - {y^2}}\gamma } \right)}\left| {{d_{{{x^2} - {y^2}} \uparrow }}{\gamma _ \uparrow }} \right\rangle  + \\\nonumber
&&L_5^{\left( {\beta d_{xy}} \right)}\left| {{\beta _ \uparrow }{d_{xy \uparrow }}} \right\rangle  + L_5^{\left( {\beta d_{xz}} \right)}\left| {{\beta _ \uparrow }{d_{xz \uparrow }}} \right\rangle  + L_5^{\left( {\beta d_{yz}} \right)}\left| {{\beta _ \uparrow }{d_{yz \uparrow }}} \right\rangle  + L_5^{\left( {\beta \gamma } \right)}\left| {{\beta _ \uparrow }{\gamma _ \uparrow }} \right\rangle 
\end{eqnarray}
In these states, the nonzero probability hole density is located on the ${d_{{x^2} - {y^2}}}$-, ${d_{xy}}$-, ${d_{xz}}$-, ${d_{yz}}$-, $\beta $-, $\gamma $-orbitals (Fig.~\ref{fig:eigenstates_str}b). The probability density distribution $\left\langle {{L_m}} \right|{N_\lambda }\left| {{L_m}} \right\rangle $, where ${N_\lambda } = \sum\limits_\sigma {a_{\lambda \sigma }^\dag {a_{\lambda \sigma }}} $ is the operator of the number of particles on the $\lambda $ orbital, in a cluster with two holes is shown in Fig.~\ref{fig:eigenstates_str}b. The second excited state is also a fourfold degenerate in energy state with the singlet and three triplet components $\left| {{L_6}} \right\rangle  - \left| {{L_9}} \right\rangle $ but with the nonzero probability density on the ${d_{{x^2} - {y^2}}}$-,${d_{{3z^2} - {r^2}}}$-, $\beta $-orbitals:
\begin{eqnarray}\label{eq:tripletT0_2exc}
&&\left| {L_6^S} \right\rangle  = L_6^{\left( {d_{{x^2} - {y^2}}d_{{3z^2} - {r^2}}} \right)}\left( {\left| {{d_{{{x^2} - {y^2}} \downarrow }}{d_{{{3z^2} - {r^2}} \uparrow }}} \right\rangle  - \left| {{d_{{{x^2} - {y^2}} \uparrow }}{d_{z \downarrow }}} \right\rangle } \right) + \\\nonumber
&& L_6^{\left( {d_{{x^2} - {y^2}}\beta } \right)}\left( {\left| {{d_{{{x^2} - {y^2}} \downarrow }}{\beta _ \uparrow }} \right\rangle  - \left| {{d_{{{x^2} - {y^2}} \uparrow }}{\beta _ \downarrow }} \right\rangle } \right) + L_6^{\left( {d_{{x^2} - {y^2}}d_{{x^2} - {y^2}}} \right)}\left| {{d_{{{x^2} - {y^2}} \downarrow }}{d_{{{x^2} - {y^2}} \uparrow }}} \right\rangle  + \\\nonumber
&& L_6^{\left( {d_{{3z^2} - {r^2}}\beta } \right)}\left( {\left| {{d_{{{3z^2} - {r^2}} \downarrow }}{\beta _ \uparrow }} \right\rangle  - \left| {{d_{{{3z^2} - {r^2}} \uparrow }}{\beta _ \downarrow }} \right\rangle } \right) + L_6^{\left( {dzdz} \right)}\left| {{d_{{{3z^2} - {r^2}} \downarrow }}{d_{{{3z^2} - {r^2}} \uparrow }}} \right\rangle  +\\\nonumber
&& L_6^{\left( {\beta \beta } \right)}\left| {{\beta _ \downarrow }{\beta _ \uparrow }} \right\rangle  
\end{eqnarray}
\begin{eqnarray}\label{eq:tripletT0_2exc}
&&\left| {L_7^{T0}} \right\rangle  = L_7^{\left( {dxdz} \right)}\left( {\left| {{d_{{{x^2} - {y^2}} \downarrow }}{d_{{{3z^2} - {r^2}} \uparrow }}} \right\rangle  + \left| {{d_{{{x^2} - {y^2}} \uparrow }}{d_{{{3z^2} - {r^2}} \downarrow }}} \right\rangle } \right) +\\\nonumber 
&&L_7^{\left( {d_{{x^2} - {y^2}}\beta } \right)}\left( {\left| {{d_{{{x^2} - {y^2}} \downarrow }}{\beta _ \uparrow }} \right\rangle  + \left| {{d_{{{x^2} - {y^2}} \uparrow }}{\beta _ \downarrow }} \right\rangle } \right) +\\\nonumber
&&L_7^{\left( {d_{{3z^2} - {r^2}}\beta } \right)}\left( {\left| {{d_{{{3z^2} - {r^2}} \downarrow }}{\beta _ \uparrow }} \right\rangle  + \left| {{d_{{{3z^2} - {r^2}} \uparrow }}{\beta _ \downarrow }} \right\rangle } \right)
\end{eqnarray}
\begin{eqnarray}\label{eq:tripletT_1_2exc}
&&\left| {L_8^{T - 1}} \right\rangle  = L_8^{\left( {d_{{x^2} - {y^2}}dz} \right)}\left| {{d_{{{x^2} - {y^2}} \downarrow }}{d_{{{3z^2} - {r^2}} \downarrow }}} \right\rangle  + L_8^{\left( {d_{{x^2} - {y^2}}\beta } \right)}\left| {{d_{{{x^2} - {y^2}} \downarrow }}{\beta _ \downarrow }} \right\rangle  +\\\nonumber
&& L_8^{\left( {d{{3z^2} - {r^2}}\beta } \right)}\left| {{d_{{{3z^2} - {r^2}} \downarrow }}{\beta _ \downarrow }} \right\rangle
\end{eqnarray}
\begin{eqnarray}\label{eq:tripletT1_2exc}
&&\left| {L_9^{T1}} \right\rangle  = L_9^{\left( {d_{{x^2} - {y^2}}d_{{3z^2} - {r^2}}} \right)}\left| {{d_{{{x^2} - {y^2}} \uparrow }}{d_{{{3z^2} - {r^2}} \uparrow }}} \right\rangle  + L_9^{\left( {d_{{x^2} - {y^2}}\beta } \right)}\left| {{d_{{{x^2} - {y^2}} \uparrow }}{\beta _ \uparrow }} \right\rangle  +\\\nonumber
&& L_9^{\left( {d_{{3z^2} - {r^2}}\beta } \right)}\left| {{d_{{{3z^2} - {r^2}} \uparrow }}{\beta _ \uparrow }} \right\rangle  
\end{eqnarray}
All two-hole states can be divided into three groups according to the distribution of the predominant number of holes  $\left\langle {{L_m}} \right|{N_\lambda }\left| {{L_m}} \right\rangle $ over $d$- and $p$-orbitals (or the probability density ${\left| {L_m^{\left( {\lambda \lambda '} \right)}} \right|^2}$ of the component $\left| {\lambda \lambda '} \right\rangle $). The first two types are the $d$- or $p$- states just like for the single-hole eigenstates. The third type of states are mixed $pd$-states with equal distribution of probability density of two holes over the $d$- and $p$-orbitals.

\section{The influence of Coulomb interaction on the energy structure of local states}
\label{sec:Coul_inter}

\begin{figure}
\centering
\includegraphics[width=0.8\linewidth]{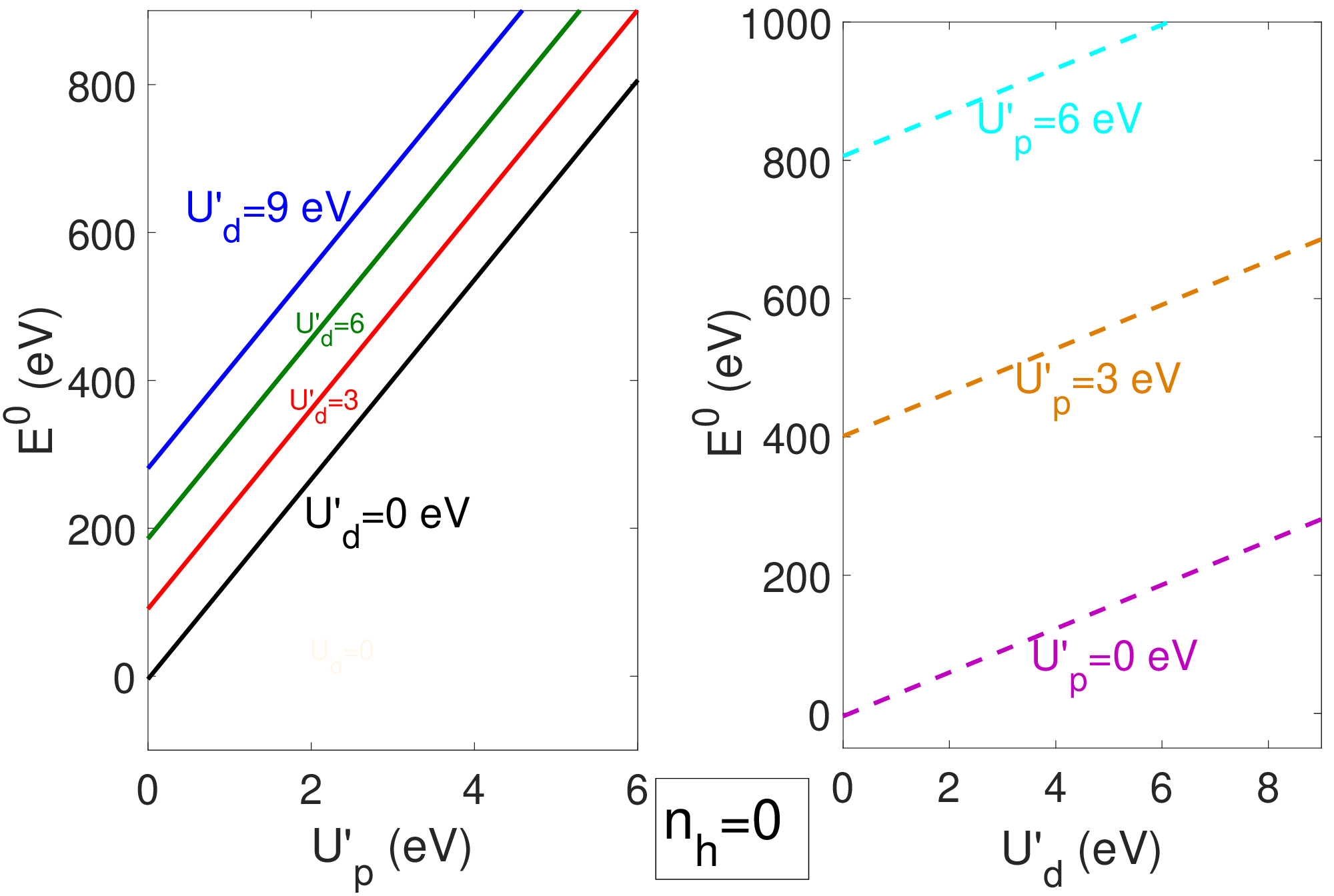}
\caption{(Color online) Dependence of the energy $E^{\left( 0 \right)}$ of the zero-hole CuO cluster eigenstate (all $d$- and $p$-orbitals are fully occupied by electrons) on the Coulomb parameters $U'_d=\left\{ {U_d, V_d, J_d} \right\}$ (left panel) and $U'_p=\left\{ {U_p, V_{pd}} \right\}$ (right panel).}
\label{fig:en_0}
\end{figure}

The Coulomb interaction leads to a change in the energy of local states and a redistribution of the probability density over different orbitals within these states. The degree of energy change and the type of redistribution of the probability density of finding a hole (charge transfer from $d$-orbitals to $p$-orbitals or vice versa) is determined by the type of Coulomb interaction. The two categories of Coulomb interactions can be distinguished depending on the type of impact on local states. The first category denoted by ${U'_d}$ includes the intraatomic Coulomb interaction on copper orbitals: the intraorbital $U_d$, the interobital $V_d$ and the Hund exchange $J_d$. Each of these three interactions increases the energy of electrons, or decrease the energy of holes, on the copper $d$-orbitals and leads to the transfer of the hole to $d$-orbitals. The second category denoted by ${U'_p}$ includes the intra-atomic Coulomb interaction on the oxygen $p$-orbitals $U_p$ and the interatomic interaction $V_{pd}$ between carriers on the copper and oxygen orbitals. The Coulomb interaction $U_p$ decreases the energy of a hole on the oxygen $p$-orbitals and results in the transfer of a hole to these $p$-orbitals. The interatomic interaction $V_{pd}$ makes a similar effect on local states as the parameter $U_p$. It is caused by the fact that each electron on the oxygen $p$-orbital interacts with $10$ electrons on the fully occupied copper $d$-shell whereas each electron on the copper $d$-orbital interacts only with $6$ electrons on the fully occupied oxygen $p$-shell. This means that the energy gain due to presence of a hole on the $p$-orbital is more than the decrease in energy due to the presence of a hole on the $d$-orbital. Therefore, we will further study the influence on local electronic states not of each Coulomb parameter separately, but of each of the two types assuming that the parameters within the two categories are related by the fixed relations: ${V_d} = \left( {{7 \mathord{\left/
 {\vphantom {7 9}} \right.
 \kern-\nulldelimiterspace} 9}} \right){U_d}$, ${J_d} = \left( {{1 \mathord{\left/
 {\vphantom {1 9}} \right.
 \kern-\nulldelimiterspace} 9}} \right){U_d}$, ${V_{pd}} = \left( {{1 \mathord{\left/
 {\vphantom {1 2}} \right.
 \kern-\nulldelimiterspace} 2}} \right){U_p}$. These relations approximately correspond to the relations chosen for HTSC cuprates based on the first-principles calculations~\cite{Gunnarsson1989,Schluter88,Schluter89,Hybertsen89,Hybertsen90,Mahan90,Grant1992,Anisimov1991,Anisimov1992}.

The dependence of the energies of local states on the magnitude of Coulomb parameters ${U'_d}$ and ${U'_p}$ is different for different numbers of holes in the CuO cluster. The strongest dependence on Coulomb parameters is observed for the energy of the local vacuum state of holes ${\varepsilon _0}$ (Fig.~\ref{fig:en_0}a,b). All copper orbitals are completely filled with electrons in this state which means that all terms of the intraorbital Coulomb repulsion will be active.

In the CuO cluster with one hole, if an electron will be missing from the orbital $d_{\zeta }$, then there will be no contribution to the energy from the Coulomb term ${U_d}d_{\zeta \downarrow }^\dag d_{\zeta \downarrow }d_{\zeta \uparrow }^\dag {d_{\zeta \uparrow }}$. For example, the term ${U_d}d_{{{x^2} - {y^2}} \downarrow }^\dag {d_{{{x^2} - {y^2}} \downarrow }}d_{{{x^2} - {y^2}} \uparrow }^\dag {d_{{{x^2} - {y^2}} \uparrow }}$ will not contribute to the energy of the ground single-hole state $\left| {{b_1}} \right\rangle $. Therefore, the dependence of the energies $E_m^{\left( 1 \right)}$ on ${U_d}$ turns out to be weaker than the dependence of the energy of the zero-hole state $E_m^{\left( 0 \right)}={\varepsilon _0}$. The difference in the hole energies on the copper and oxygen orbitals increases with increasing ${U_d}$. As a result, the contribution of the oxygen orbitals to the low-energy single-hole states decreases. At large ${U_d}$, this contribution tends to zero, and the difference between the energies of the single-hole states $E_m^{\left( 1 \right)}$ with $m=1-5$ having the pure nature of the $d$-orbitals becomes the same as between the energies of $d$-orbitals in a isolated copper atom.
\begin{figure}
\centering
\includegraphics[width=0.8\linewidth]{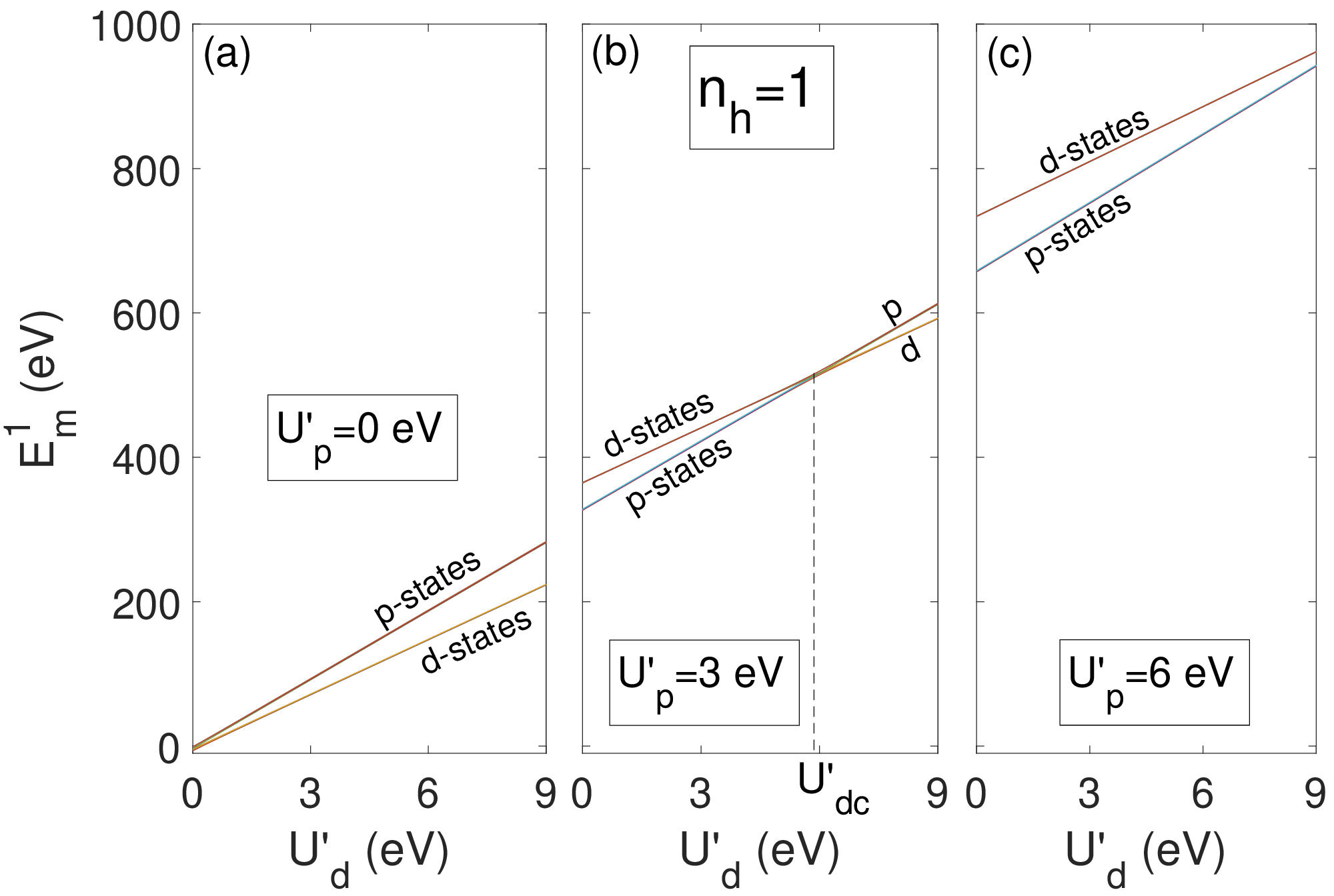}
\caption{(Color online) Dependence of the single-hole CuO cluster eigenstates energies $E_m^{\left( 1 \right)}$ ($m=1-16$) on the Coulomb parameter $U'_d=\left\{ {U_d, V_d, J_d} \right\}$ at fixed $U'_p=\left\{ {U_p, V_{pd}} \right\}$. The notations $d$ ($p$) near line segments denote that the energy in this segment is related to the $d$-states ($p$-states) with predominant probability density on the $d$($p$)-orbitals. Each line is a bundle of dependencies with the same type of eigenstates ($d$ or $p$).}
\label{fig:en_1_Ud}
\end{figure}
\begin{figure}
\centering
\includegraphics[width=0.8\linewidth]{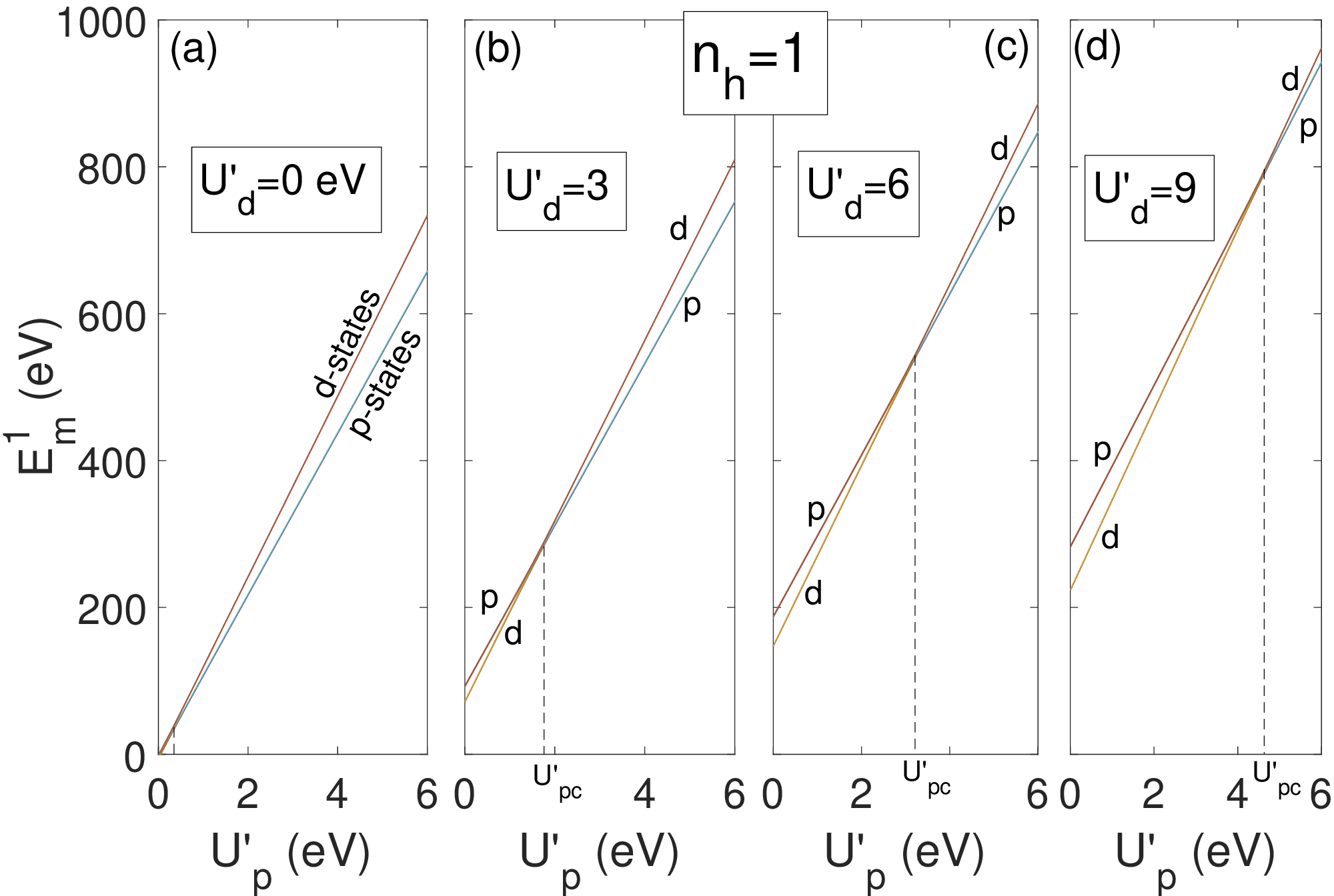}
\caption{(Color online) Same as in Fig.~\ref{fig:en_1_Ud}, but it is the dependence on the Coulomb parameter $U'_p=\left\{ {U_p, V_{pd}} \right\}$ at fixed $U'_d=\left\{ {U_d, V_d, J_d} \right\}$.}
\label{fig:en_1_Up}
\end{figure}

\begin{figure}
\centering
\includegraphics[width=0.8\linewidth]{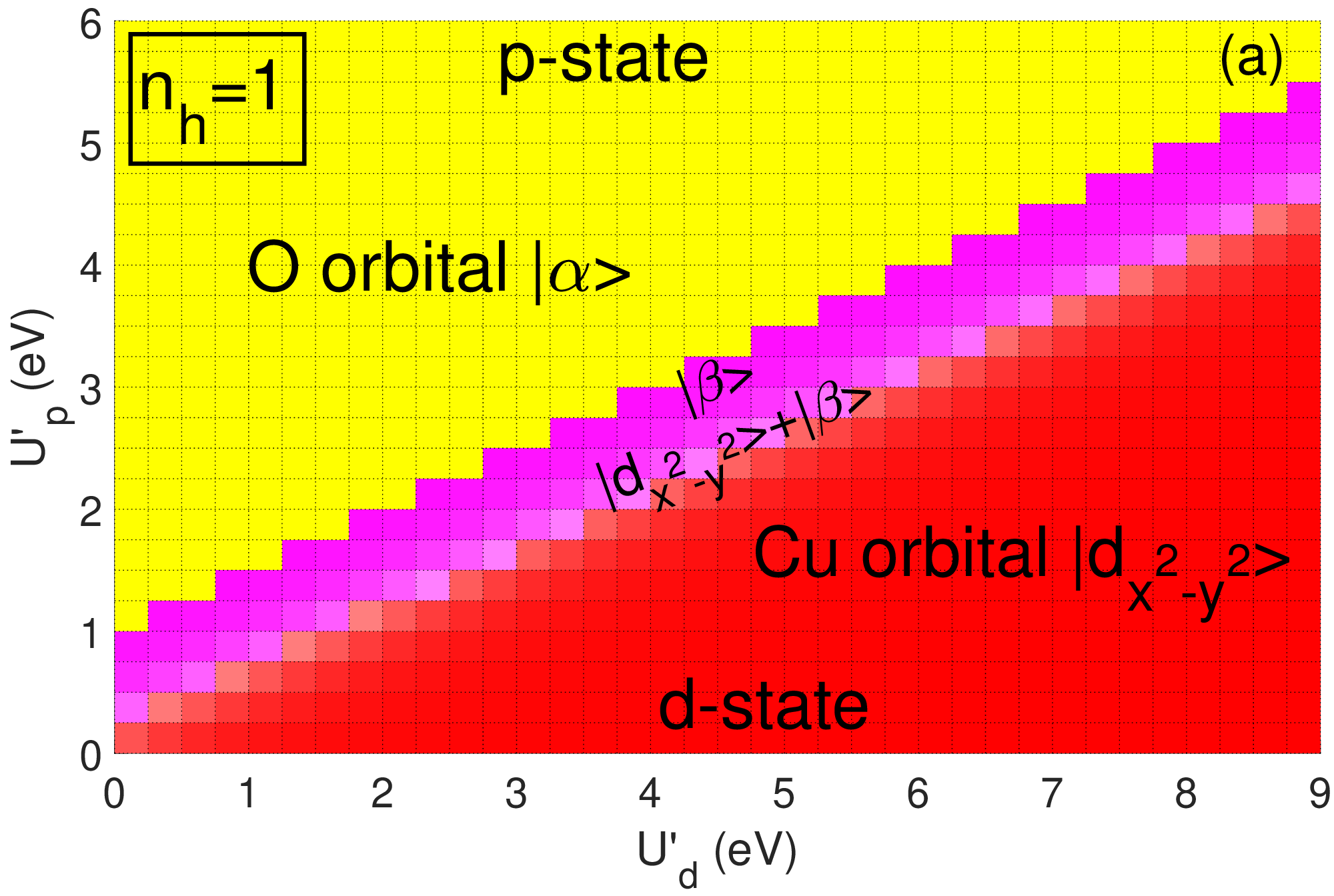}
\includegraphics[width=0.8\linewidth]{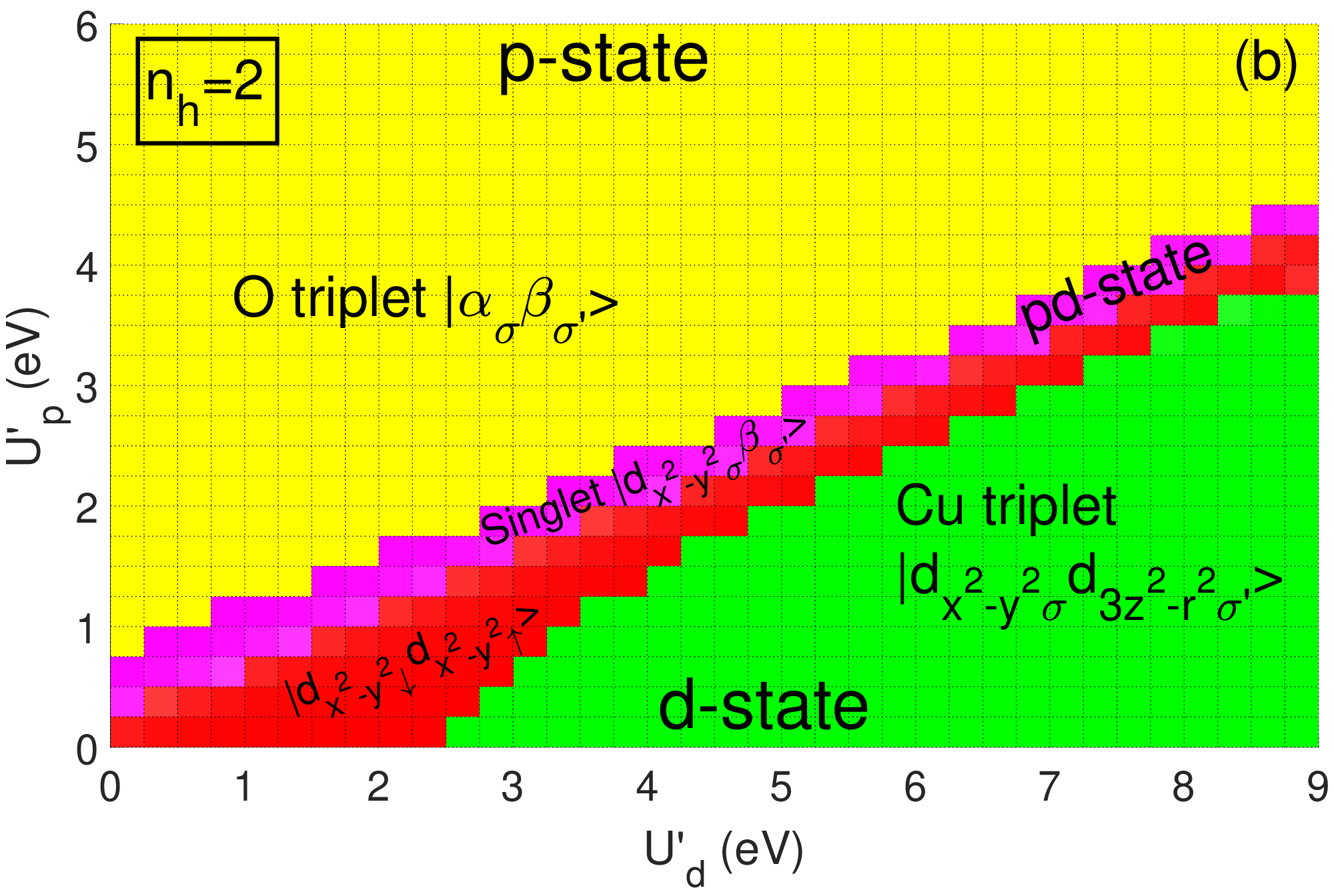}
\caption{(Color online) (a) The orbital $\left| \lambda  \right\rangle $ of the ground single-hole eigenstate and (b) the component $\left| {\lambda \lambda '} \right\rangle $ of the ground two-hole eigenstate with maximal probability density ${\left| {c_g^{\left( \lambda  \right)}} \right|^2}$ and ${\left| {L_g^{\left( {\lambda \lambda '} \right)}} \right|^2}$, respectively, in the space of Coulomb parameters $U'_d=\left\{ {U_d, V_d, J_d} \right\}$ and $U'_p=\left\{ {U_p, V_{pd}} \right\}$.}
\label{fig:phdiag_1_2}
\end{figure}

\begin{figure}
\centering
\includegraphics[width=0.8\linewidth]{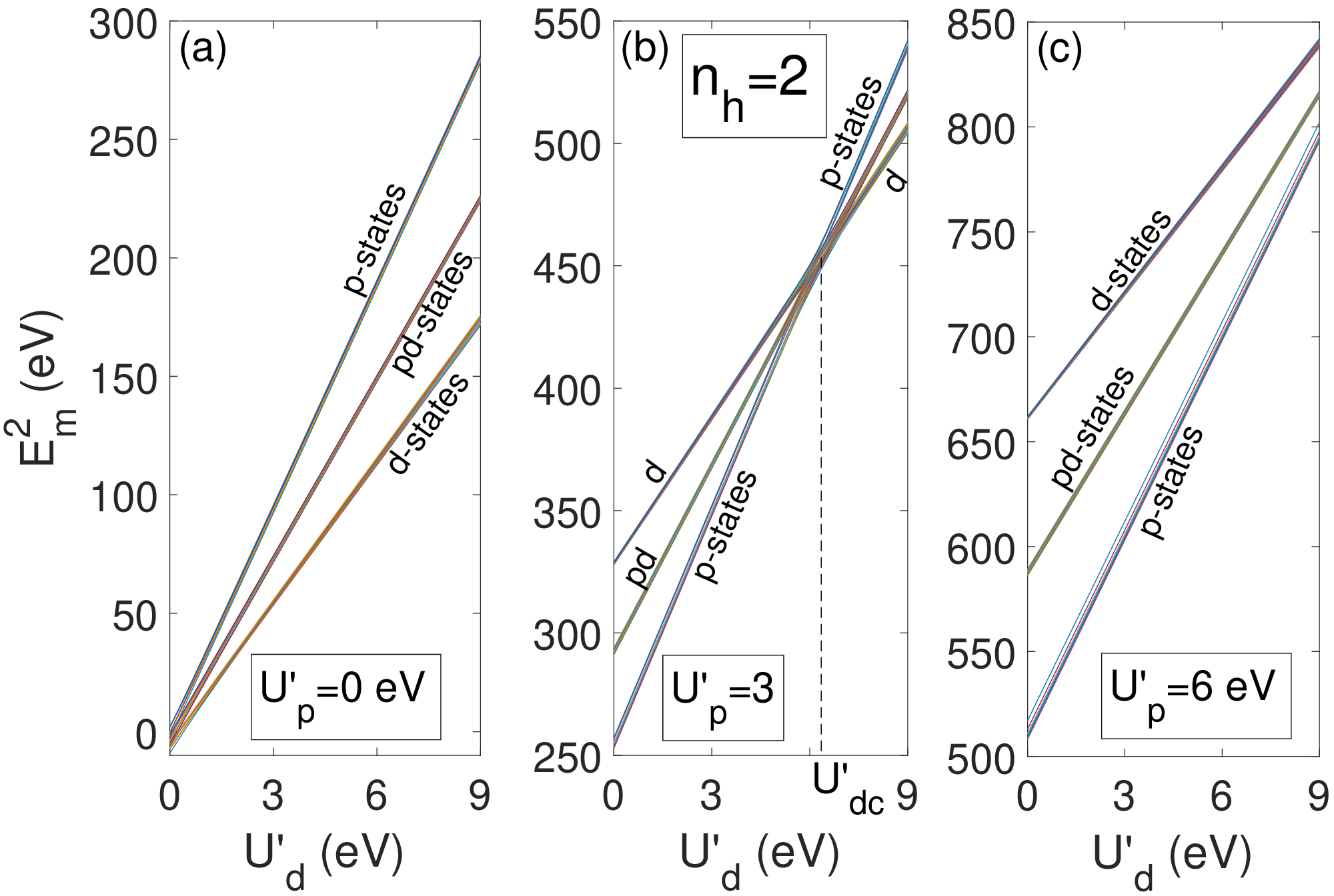}
\caption{(Color online) Dependence of the two-hole CuO cluster eigenstates energies $E_m^{\left( 2 \right)}$ ($m=1-120$) on the Coulomb parameter $U'_d=\left\{ {U_d, V_d, J_d} \right\}$ at fixed $U'_p=\left\{ {U_p, V_{pd}} \right\}$. The notations $d$, $p$ and $pd$ near line segments denote that the energy in this segment is related to the $d$-, $p$- and  $pd$-states with predominant probability density on the $d$-, $p$- and both orbitals, respectively. Each line is a bundle of dependencies with the same type of states ($d$ or $p$).}
\label{fig:en_2_Ud}
\end{figure}

\begin{figure}
\centering
\includegraphics[width=0.8\linewidth]{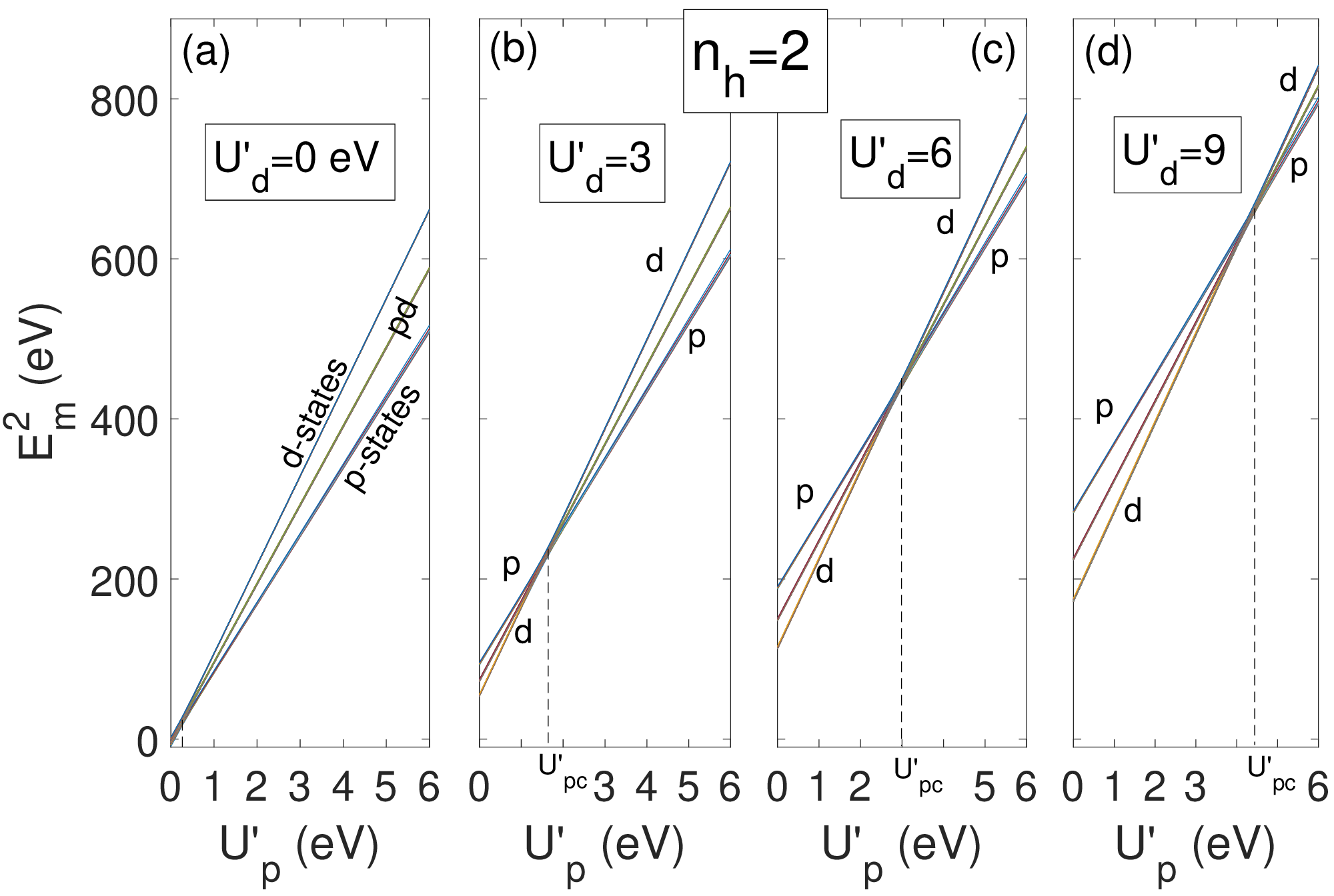}
\caption{(Color online) Same as in Fig.~\ref{fig:en_2_Ud}, but it is the dependence on the Coulomb parameter $U'_p=\left\{ {U_p, V_{pd}} \right\}$ at fixed $U'_d=\left\{ {U_d, V_d, J_d} \right\}$.}
\label{fig:en_2_Up}
\end{figure}

Since there are two types of single-hole states, there are correspondingly two types of dependencies on ${U'_d}$ or ${U'_p}$ (Figs.~\ref{fig:en_1_Ud}a-c). Each thick line in Figs.~\ref{fig:en_1_Ud}a-c, Figs.~\ref{fig:en_1_Up}a-d, Figs.~\ref{fig:en_2_Ud}a-c, Figs.~\ref{fig:en_2_Up}a-d is actually a bundle of lines corresponding to different eigenstates that are indistinguishable on a large energy scale. Thus, the $d$-state line in Figs.~\ref{fig:en_1_Ud}a-c and Figs.~\ref{fig:en_1_Up}a-d includes lines of the five eigenstates $\left| {{b_m}} \right\rangle $ with $m=1-5$, and the $p$-state line consists of lines of the three $\left| {{b_m}} \right\rangle $ with $m=6-8$. It is seen from Fig.~\ref{fig:en_1_Ud}a that at ${U'_p}=0$ eV, the energies of the single-hole states $\left| {{b_m}} \right\rangle $ with $m=1-5$ mostly formed by $d$-orbitals monotonically increase with increasing ${U'_d}$, the state $\left| {{b_1}} \right\rangle $ remains the ground one throughout the entire range of ${U'_d}$ variation. The energies of states $\left| {{b_m}} \right\rangle $ with $m=6-8$ having the character of the $p$-orbitals increase faster with increasing ${U'_d}$ than the energies of the $d$-states. It is caused by the fact that the copper orbitals in the single-hole $d$-states are less filled with electrons than the $p$-states and the energy of the Coulomb interactions ${U'_d}$ for them is lower. Therefore, at ${U'_p}=0$ eV, the $p$-state $\left| {{b_1}} \right\rangle $ never crosses the ground state $\left| {{b_1}} \right\rangle $ with increasing ${U'_d}$. Almost the entire probability density of a holes goes to the $d_{{x^2} - {y^2}}$-orbital with further increasing ${U'_d}$ (Fig.~\ref{fig:phdiag_1_2}a red region). At large values of the parameter ${U'_p}$, each of the dependencies $E_m^1\left( {{{U'}_d}} \right)$ preserve their slopes and each of these is transferred to high energies as a whole. The oxygen orbitals in the single-hole $d$-states are significantly more filled with electrons and therefore the energy of the Coulomb interaction ${U'_p}$ for them is higher. Therefore, the amount of energy by which the $d$-states rise is greater than that by which the $p$-states rise. The crossover between $d$- and $p$-states appears at ${U'_{d}}={U'_{dc}}$ (Fig.~\ref{fig:en_1_Ud}b). The larger ${U'_p}$, the larger ${U'_{dc}}$. The ground state at ${U'_d}<{U'_{dc}}$ is the eigenstate $\left| {{b_1}} \right\rangle $ (a hole is mainly on $\alpha$-orbital, Fig.~\ref{fig:phdiag_1_2}a, yellow region). At ${U'_d}>{U'_{dc}}$, the probability density goes from the oxygen $\alpha$-orbital to the copper $d_{{x^2} - {y^2}}$-orbital with a small mixture of the oxygen $\beta$-orbital (Fig.~\ref{fig:phdiag_1_2}a, red region). At ${U'_p}=6$ eV, the crossover point is at ${U'_d}>9$ eV (Fig.~\ref{fig:en_1_Ud}c). Therefore the ground state has the character of the oxygen $\alpha$-orbital at the whole range ${U'_d}$ from $0$ to $9$ eV.  

The dependencies of the energy of the single-hole $d$- and $p$-states on ${U'_p}$ with fixed ${U'_d}$ also have different slopes (Figs.~\ref{fig:en_1_Up}a-d). The energies of the $p$-states increase slower with increasing ${U'_p}$ than the energies of the $d$-states. Therefore, it is obvious that the crossover of the $d$- and $p$-states also occurs in the case of the ${U'_p}$ variation. The crossover point ${U'_{pc}}$ also shifts to larger values with increasing ${U'_d}$ (Figs.~\ref{fig:en_1_Up}a-d). The ground state at ${U'_p}<{U'_{pc}}$ is the eigenstate $\left| {{b_1}} \right\rangle $ (the hole is mainly on the copper $d_{{x^2} - {y^2}}$-orbital with a small mixture of the oxygen $\beta$-orbital, Fig.~\ref{fig:phdiag_1_2}a, red region). At ${U'_p}>{U'_{pc}}$, the probability density goes to the oxygen $\alpha$-orbital of the eigenstate $\left| {{b_1}} \right\rangle $. Thus, there are two large regions of parameters ${U'_d}$-${U'_p}$ in which the ground single-hole state has either the nature of a copper $d_{{x^2} - {y^2}}$-orbital (Fig.~\ref{fig:phdiag_1_2}a, red region) or the nature of the oxygen $\alpha$-orbital (Fig.~\ref{fig:phdiag_1_2}a, yellow region). In the parameter space $U'_d$-$U'_p$, the crossover points form a line. The redistribution of the probability density between the $d$- and $p$-orbitals when moving from one region to another occurs smoothly across a boundary region (Fig.~\ref{fig:phdiag_1_2}, purple region). At Coulomb parameters within this boundary region, the ground single-hole eigenstate is a mixed state with approximately equal probability density on the $d_{{x^2} - {y^2}}$- and $\beta$-orbitals or with a predominance of the $\beta$-orbital (Fig.~\ref{fig:phdiag_1_2}, purple region).   
 
Despite the fact that the distribution of two holes by eight orbitals has a much more complex form than the distribution of one hole, the influence of the Coulomb interactions on the energies of the two-hole states is similar to that for the single-hole states. There is three types of dependencies of energy on Coulomb parameters (Figs.~\ref{fig:en_2_Ud}a-c, Figs.~\ref{fig:en_2_Up}a-d). The energy of the $pd$-states is between the energies of the d- and $p$-states. The behavior of the energies of the two-hole $d$- and $p$-states with increasing ${U'_d}$ and ${U'_p}$ repeats behavior of the analogous types of the single-hole states. The crossover in the two-hole Hilbert space sector occurs between the triplet $p$-state with the components $\left| {{\alpha _ \downarrow }{\beta _ \downarrow }} \right\rangle $, $\left| {{\alpha _ \uparrow }{\beta _ \uparrow }} \right\rangle $, $\frac{1}{{\sqrt 2 }}\left( {\left| {{\alpha _ \downarrow }{\beta _ \uparrow }} \right\rangle  + \left| {{\alpha _ \uparrow }{\beta _ \downarrow }} \right\rangle } \right)$ (Fig.~\ref{fig:phdiag_1_2}b, yellow region) and the triplet $d$-state with the components $\left| {{d_{{x^2} - {y^2} \downarrow }}{d_{3{z^2} - {r^2} \downarrow }}} \right\rangle $, $\left| {{d_{{x^2} - {y^2} \uparrow }}{d_{3{z^2} - {r^2} \uparrow }}} \right\rangle $, $\frac{1}{{\sqrt 2 }}\left( {\left| {{d_{{x^2} - {y^2} \downarrow }}{d_{3{z^2} - {r^2} \uparrow }}} \right\rangle  + \left| {{d_{{x^2} - {y^2} \uparrow }}{d_{3{z^2} - {r^2} \downarrow }}} \right\rangle } \right)$ (Fig.~\ref{fig:phdiag_1_2}b, red region). The $pd$-state in the form of the Zhang-Rice singlet becomes the ground state forming the boundary region near the crossover line (Fig.~\ref{fig:phdiag_1_2}b, purple region) between large regions of the $d$- and $p$-states.

The regions of $d$- and $p$-states and the boundary region of $pd$ states almost coincide for the ground single- and two-hole eigenstates (Fig.~\ref{fig:phdiag_1_2}a,b). The transition between the ground single- and two-hole eigenstates defines the quasiparticle excitation $\left( {{b_g}{L_g}} \right)$, where $\left| {{b_g}} \right\rangle $ is the ground single-hole eigenstates and $\left| {{L_g}} \right\rangle $ is the ground two-hole eigenstate, which forms the top of the valence band. The orbital structure of the initial and final states of the excitation determines the orbital character of this quasiparticle. Therefore, orbital character of the quasiparticle $\left( {{b_g}{L_g}} \right)$ will differ in the the different regions of the parameter space ${U'_d}$-${U'_p}$.

\section{The gap between the energies of dispersionless quasiparticles forming conductivity and valence bands}
\label{sec:quasiparticles}
The energies of local quasiparticle excitations and their ratio are important to analyze the electronic structure of quasiparticle excitations and to estimate the magnitude of the band gap. Excitations between zero-hole and single-hole cluster eigenstates form the upper Hubbard band which is the conductivity band. The excitations between the single-hole and two-hole states form the lower Hubbard band which is the valence band. The region of low-energy excitations near Fermi level is formed by the excitation $\left( {0{b_g}} \right)$ with the energy $\Omega_{01}$ between the zero-hole $\left| 0 \right\rangle $ and ground single-hole eigenstate $\left| {{b_g}} \right\rangle $ (Fig.~\ref{fig:levels}, blue dashed curve), where $b_g$ changes depending on Coulomb or the parameters, and the excitation $\left( {{b_g}{L_g}} \right)$ with the energy $\Omega_{12}$ (Fig.~\ref{fig:levels}, green dashed curve) between $\left| {{b_g}} \right\rangle $ and the ground two-hole eigenstate $\left| {{L_g}} \right\rangle $. The excitation $\left( {{b_g}{L_g}} \right)$  forms the top of the valence band. The energy of these two quasiparticle excitations are determined as ${\Omega _{01}} = E_g^{\left( 1 \right)} - {E^{\left( 0 \right)}}$ and ${\Omega _{12}} = E_g^{\left( 2 \right)} - E_g^{\left( 1 \right)}$~\cite{Ovchinnikov89,Gavrichkov00,Korshunov05,OvchinnikovValkov}, where $E_g^{\left( 1 \right)}$ and $E_g^{\left( 2 \right)}$ are the energies of the ground single- and two-hole states. The band gap is the gap between the bands of the quasiparticles $\left( {0{b_g}} \right)$ and $\left( {{b_g}{L_g}} \right)$ with taking into account their dispersion. However, difference between dispersionless quasiparticles energies $\Delta={\Omega _{12}-{\Omega _{01}}}$ can be used to estimate the possibility formation of the insulating or metallic state.

\begin{figure}
\centering
\includegraphics[width=0.9\linewidth]{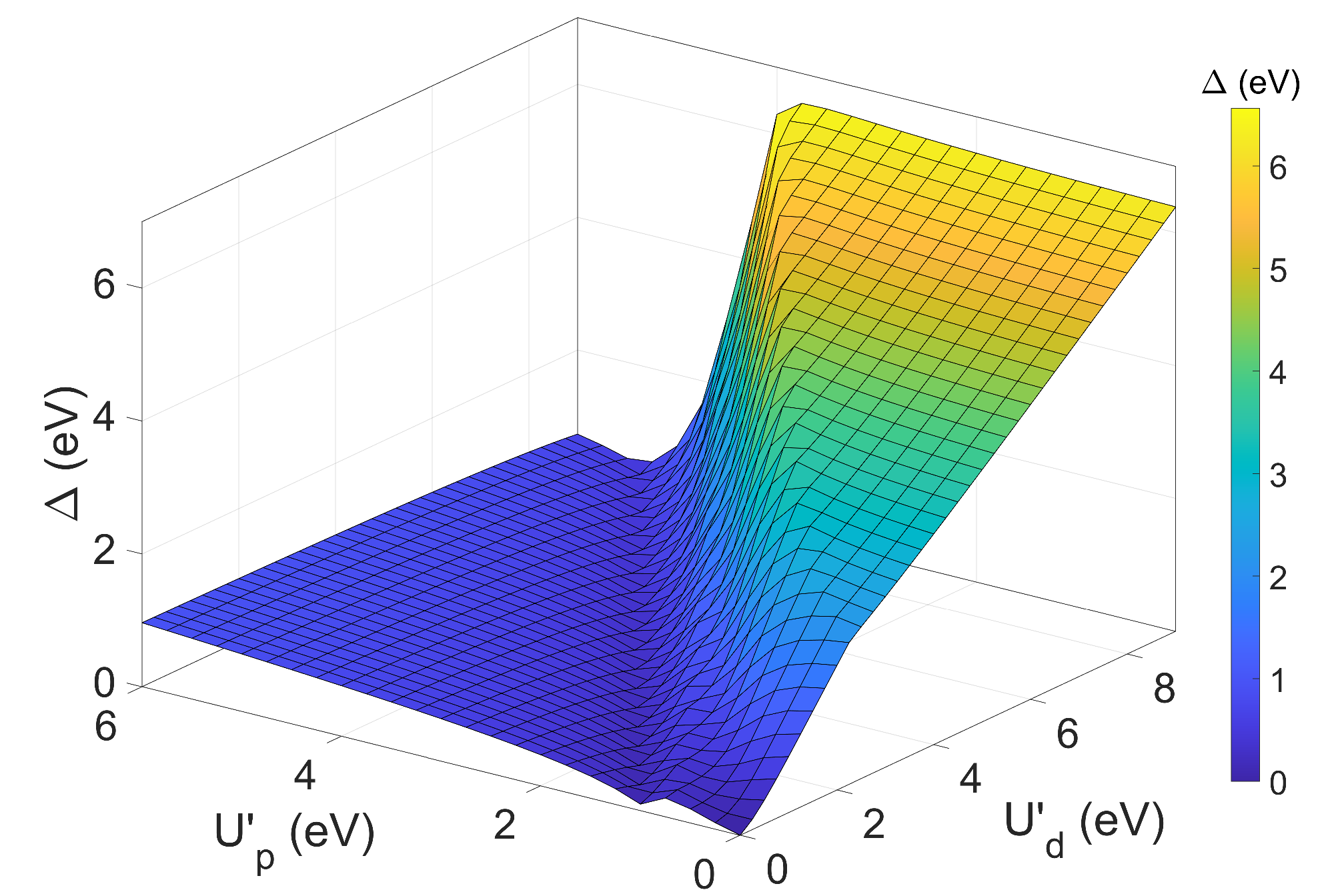}
\caption{(Color online) The value of the gap $\Delta={\Omega _{12}-{\Omega _{01}}}$ between the dispersionless quasiparticles $\left( {0{b_g}} \right)$ and $\left( {{b_g}{L_g}} \right)$ with the energies ${\Omega _{01}}=E_g^{\left( 1 \right)} - {E^{\left( 0 \right)}}$ and $\Omega _{12}=E_g^{\left( 2 \right)} - E_g^{\left( 1 \right)}$ where ${E^{\left( 0 \right)}}$, $E_g^{\left( 1 \right)}$, $E_g^{\left( 2 \right)}$ are the energies of the ground zero-, single- and two-hole CuO cluster eigenstates, in the space of the Coulomb parameters $U'_d=\left\{ {U_d, V_d, J_d} \right\}$ and $U'_p=\left\{ {U_p, V_{pd}} \right\}$.}
\label{fig:gap}
\end{figure}
The value of the gap $\Delta$ in the space of the Coulomb parameters $U'_d$ and $U'_p$ is shown in Fig.~\ref{fig:gap}. The correspondence between the map of the $\Delta$ value (Fig.~\ref{fig:gap}) and the maps of the maximum probability densities of a hole on various orbitals in the ground multihole states (Figs.~\ref{fig:phdiag_1_2}a,b) is visible. The gap between the dispersionless quasiparticles on the top of the valence band and the conductivity band have small values ($\sim1$ eV) in the same region (Fig.~\ref{fig:gap}) where the ground state is filled with oxygen holes (Fig.~\ref{fig:phdiag_1_2}a,b, yellow region). At large $U'_d$ and small $U'_p$ (the region of $d$-states in Figs.~\ref{fig:phdiag_1_2}a,b, red region), the $\Delta$ magnitude is large ($\sim6$ eV). The intermediate region of Coulomb parameters at which a sharp increase in the gap size occurs (Fig.~\ref{fig:gap}, region of steep slope) corresponds to the boundary region between $d$- and $p$-states (Figs.~\ref{fig:phdiag_1_2}a,b, purple region). In the intermediate region, the insignificant variation of $U'_d$ or $U'_p$ leads to the radical change in the orbital character of the local ground multiparticle states and the gap $\Delta$ magnitude. The smallest gap size is observed at Coulomb parameters located on the boundary region, $\Delta$ is about $0.5$ eV. Since the width of the bands after taking into account quasiparticle dispersion can exceed this value, it can be expected that a metallic state can be realized at the parameters in the boundary region, and also, possibly, in the region of the predominance of the $p$-states. At the Coulomb parameters which corresponds to their values in cuprates ($U'_d=9$, $U'_p=4$), the ground state of the electronic system of the CuO cluster is in $d$-state near the boundary region, the ground single- and two-hole states are the states with holes on the $d_{{x^2} - {y^2}}$-orbitals and $\Delta$ is approximately $5$ eV. However, we can get into the boundary region with a slight change by approximately $0.5$ eV in one of the parameters $U'_d$ or $U'_p$ , where the characteristics of the system are very similar to cuprates: the orbital character of local states is represented by the mixture of $d_{{x^2} - {y^2}}$- and $\beta$-orbitals and $\Delta=2-3$ eV. 

\section{Conclusion}\label{sec:conclusion}
Here we calculated the eigenstates and eigenenergies of the CuO cluster with zero-, one and two holes for different values of Coulomb parameters. The single- and two-hole eigenstates can be divided into $d$-states in which holes are predominantly in $d$-orbitals and $p$-states in which the $p$-orbitals have the predominant probability density of holes. For the two-hole states, there are also $pd$-states with a uniform distribution of holes over $d$- and $p$-orbitals. Each of these types of states has different dependencies of energy on the Coulomb parameters, therefore, at some values of the latter, a crossover between $d$- and $p$-states occurs. The crossover line divides the space of Coulomb parameters into the two large regions in which the ground single- and two-hole states have the nature of $d$- or $p$-orbitals and the magnitude of the gap $\Delta$ between the dispersionless quasiparticle excitations that form the conduction band and the top of the valence band differs strongly. In the region of $p$-states, $\Delta$ has small values, $\Delta$ in the region of $d$-states is significantly larger. The states in the boundary region near the crossover line have a mixed nature of the $d$- and $p$-orbitals, $\Delta$ value changes sharply in a narrow range of variation of the Coulomb parameters. Knowledge of the orbital and energy structure of the local multiparticle states serves as a basis for understanding the nature of quasiparticles in the CuO monolayer and the band gap magnitude. It is clear from the obtained results that the band gap magnitude and the orbital nature of quasiparticle excitations at the top of the valence band and at the conductivity band will depend strongly on the value of the Coulomb parameters.

\section*{Acknowledgments}
%\begin{acknowledgments}
We are grateful to L.V. Begunovich and I.A. Nekrasov for useful discussions. The present work was carried out within the state assignment of Kirensky Institute of Physics.
%\end{acknowledgments}

%\begin{table}[t]%% placement specifier
%%% Use tabular environment to tag the tabular data.
%%% https://en.wikibooks.org/wiki/LaTeX/Tables#The_tabular_environment
%\centering%% For centre alignment of tabular.
%\begin{tabular}{l c r}%% Table column specifiers
%%% Tabular cells are separated by &
%  1 & 2 & 3 \\ %% A tabular row ends with \\
%  4 & 5 & 6 \\
%  7 & 8 & 9 \\
%\end{tabular}
%%% Use \caption command for table caption and label.
%\caption{Table Caption}\label{fig1}
%\end{table}

%% Use figure environment to create figures
%% Refer following link for more details.
%% https://en.wikibooks.org/wiki/LaTeX/Floats,_Figures_and_Captions

\appendix
\section{The values of electron on-site energies and hopping integrals}\label{app:parameters}
The values of on-site energies and hopping integrals of eight-band $p-d$ model Hamiltonian (\ref{eq:Ham_el}) are given in the Tables~\ref{table1}-\ref{table4}.
\begin{table}
\centering
\begin{tabular}{|c|c|c|c|c|c|}
\hline
\rule{0mm}{5mm} ${\varepsilon _{{{\rm{d}}_{{x^2} - {y^2}}}}}$ & ${\varepsilon _{d_{3{z^2} - {r^2}}}}$ & ${\varepsilon _{d_{xy}}}$ & ${\varepsilon _{d_{xz}}}$ (${\varepsilon _{d_{yz}}}$) & ${\varepsilon _{p_x}}$ (${\varepsilon _{p_y}}$) & ${\varepsilon _{p_z}}$ \\
\hline
\rule{0mm}{5mm}
$0.787$ & $0.498$ & $0.024$ & $0.3830$ & $-1.942$ & $-0.168$ \\
\hline
\end{tabular}
\caption{On-site energies of an electron on the copper $d$- and oxygen $p$-orbitals in the eight-band $p-d$ model Hamiltonian obtained within the DFT approach~\cite{Slobodchikov2023} (in units of eV)}\label{table1}
\end{table}

\begin{table}
\centering
\begin{tabular}{|c|c|c|c|c|c|}
\hline
\rule{0mm}{5mm}
${\bf{R}} = \left( {{R_x},{R_y}} \right)$ & $\left( { \frac{a}{2},\frac{b}{2}} \right)$ & $\left( { - \frac{a}{2},\frac{b}{2}} \right)$ & $\left( { \pm \frac{{3a}}{2}, \pm \frac{b}{2}} \right)$ & $\left( { \pm \frac{{3a}}{2}, \pm \frac{{3b}}{2}} \right)$ & $\left( { \pm \frac{{5a}}{2}, \pm \frac{b}{2}} \right)$ \\
& $\left( { - \frac{a}{2}, - \frac{b}{2}} \right)$ & $\left( {\frac{a}{2}, - \frac{b}{2}} \right)$ & $\left( { \pm \frac{a}{2}, \pm \frac{{3b}}{2}} \right)$ & & $\left( { \pm \frac{a}{2}, \pm \frac{{5b}}{2}} \right)$ \\
 \hline
$\left| {{t_{d_{{x^2} - {y^2}}p_x}}\left( {\bf{R}} \right)} \right|$ & $1.349$ & $0$ & $0.04$ & $0.017$ & $0.004$\\
$\left| {{t_{d_{{x^2} - {y^2}}p_y}}\left( {\bf{R}} \right)} \right|$ & $0$ & $1.349$ & $0.04$ & $0$ & $0.004$\\
$\left| {{t_{d_{3{z^2} - {r^2}}p_x}}\left( {\bf{R}} \right)} \right|$ & $0.367$ & $0$ & $0.018$ & $0.003$ & $0.002$\\
$\left| {{t_{d_{3{z^2} - {r^2}}p_y}}\left( {\bf{R}} \right)} \right|$ & $0$ & $0.367$ & $0.018$ & $0$ & $0.002$\\
$\left| {{t_{d_{xy}p_x}}\left( {\bf{R}} \right)} \right|$ & $0$ & $0.782$ & $0.029$ & $0$ & $0.003$\\
$\left| {{t_{d_{xy}p_y}}\left( {\bf{R}} \right)} \right|$ & $0.782$ & $0$ & $0.006$ & $0.01$ & $0.003$\\
$\left| {{t_{d_{xz}p_z}}\left( {\bf{R}} \right)} \right|$ & $0.784$ & $0$ & $0.012$ & $0.01$ & $0.005$\\
$\left| {{t_{d_{yz}p_z}}\left( {\bf{R}} \right)} \right|$ & $0$ & $0.784$ & $0.012$ & $0$ & $0.005$\\
\hline
$\left| {{t_{d_{{x^2} - {y^2}}p_z}}\left( {\bf{R}} \right)} \right|$ & & & & &\\
$\left| {{t_{d_{3{z^2} - {r^2}}p_z}}\left( {\bf{R}} \right)} \right|$ & & & & &\\
$\left| {{t_{d_{xy}p_z}}\left( {\bf{R}} \right)} \right|$ & & & & &\\
 $\left| {{t_{d_{xz}p_x}}\left( {\bf{R}} \right)} \right|$ & $0$ & $0$ & $0$ & $0$ & $0$\\
 $\left| {{t_{d_{yz}p_x}}\left( {\bf{R}} \right)} \right|$ & & & & &\\
  $\left| {{t_{d_{xz}p_y}}\left( {\bf{R}} \right)} \right|$ & & & & &\\
  $\left| {{t_{d_{yz}p_y}}\left( {\bf{R}} \right)} \right|$ & & & & & \\
\hline
\end{tabular}
\caption{The absolute values of the hopping integrals between the orbitals of the copper atom at the site ${\bf{f}} \equiv \left( {{f_x},{f_y}} \right)$ and the oxygen atom at the site ${\bf{f}} + {\bf{R}} = \left( {{f_x} + {R_x},{f_y} + {R_y}} \right)$ in the eight-band $p-d$ model Hamiltonian (\ref{eq:Ham_el})}\label{table2}
\end{table}

\begin{table}
\begin{tabular}{|c|c|c|c|c|c|}
\hline
\rule{0mm}{5mm}
${\bf{R}} = \left( {{R_x},{R_y}} \right)$ & $\left( { \pm a,0} \right)$ & $\left( { \pm a, \pm b} \right)$ & $\left( { \pm 2a,0} \right)$ & $\left( { \pm 2a, \pm b} \right)$ & $\left( { \pm 2a, \pm 2b} \right)$ \\
& $\left( {0, \pm b} \right)$ & & $\left( {0, \pm 2b} \right)$ & $\left( { \pm a, \pm 2b} \right)$ &\\
 \hline
$\left| {{t_{p_xp_y}}\left( {\bf{R}} \right)} \right|$ & $0.609$ & $0$ & $0.026$ & $0.001$ & $0$\\
$\left| {{t_{p_xp_x}}\left( {\bf{R}} \right)} \right|$, $\left| {{t_{p_yp_y}}\left( {\bf{R}} \right)} \right|$ & $0.353$ & $0.148$ & $0.036$ & $0.008$ & $0.001$\\
$\left| {{t_{p_zp_z}}\left( {\bf{R}} \right)} \right|$ & $0.253$ & $0.069$ & $0.008$ & $0.005$ & $0.002$\\
\hline
\end{tabular}
\caption{The absolute values of the hopping integrals between the orbitals of oxygen atoms at the sites ${\bf{g}} \equiv \left( {{g_x},{g_y}} \right)$ and ${\bf{g}} + {\bf{R}} = \left( {{g_x} + {R_x},{g_y} + {R_y}} \right)$ in the eight-band $p-d$ model Hamiltonian }\label{table3}
\end{table}

\begin{table}
\begin{tabular}{|c|c|c|c|c|c|}
\hline
\rule{0mm}{5mm}
${\bf{R}} = \left( {{R_x},{R_y}} \right)$ & $\left( { \pm a,0} \right)$ & $\left( { \pm a, \pm b} \right)$ & $\left( { \pm 2a,0} \right)$ & $\left( { \pm 2a, \pm b} \right)$ & $\left( { \pm 2a, \pm 2b} \right)$ \\
& $\left( {0, \pm b} \right)$ & & $\left( {0, \pm 2b} \right)$ & $\left( { \pm a, \pm 2b} \right)$ &\\
 \hline
$\left| {{t_{d_{{x^2} - {y^2}}d_{{x^2} - {y^2}}}}\left( {\bf{R}} \right)} \right|$ & $0.157$ & $0.01$ & $0.017$ & $0.002$ & $0.003$\\
$\left| {{t_{d_{{x^2} - {y^2}}d_{3{z^2} - {r^2}}}}\left( {\bf{R}} \right)} \right|$ & $0$ & $0.003$ & $0$ & $0.003$ & $0$\\
$\left| {{t_{d_{{x^2} - {y^2}}d_{xy}}}\left( {\bf{R}} \right)} \right|$ & $0$ & $0$ & $0$ & $0.003$ & $0$\\
$\left| {{t_{d_{{x^2} - {y^2}}d_{xz}}}\left( {\bf{R}} \right)} \right|$, $\left| {{t_{d_{{x^2} - {y^2}}d_{yz}}}\left( {\bf{R}} \right)} \right|$ & $0$ & $0$ & $0$ & $0$ & $0$\\
$\left| {{t_{d_{3{z^2} - {r^2}}d_{3{z^2} - {r^2}}}}\left( {\bf{R}} \right)} \right|$ & $0.049$ & $0.013$ & $0.02$ & $0.001$ & $0$\\
$\left| {{t_{d_{3{z^2} - {r^2}}d_{xy}}}\left( {\bf{R}} \right)} \right|$ & $0.01$ & $0$ & $0.002$ & $0$ & $0$\\
$\left| {{t_{d_{3{z^2} - {r^2}}d_{xz}}}\left( {\bf{R}} \right)} \right|$, $\left| {{t_{d_{3{z^2} - {r^2}}d_{yz}}}\left( {\bf{R}} \right)} \right|$ & $0$ & $0$ & $0$ & $0$ & $0$\\
$\left| {{t_{d_{xy}d_{xy}}}\left( {\bf{R}} \right)} \right|$ & $0.243$ & $0.04$ & $0.003$ & $0.002$ & $0.001$\\
$\left| {{t_{d_{xy}d_{xz}}}\left( {\bf{R}} \right)} \right|$, $\left| {{t_{d_{xy}d_{yz}}}\left( {\bf{R}} \right)} \right|$ & $0$ & $0$ & $0$ & $0$ & $0$\\
$\left| {{t_{d_{xz}d_{xz}}}\left( {\bf{R}} \right)} \right|$, $\left| {{t_{d_{yz}d_{yz}}}\left( {\bf{R}} \right)} \right|$ & $0.074$ & $0.051$ & $0.02$ & $0$ & $0$\\
$\left| {{t_{d_{xz}d_{yz}}}\left( {\bf{R}} \right)} \right|$ & $0.04$ & $0$ & $0.003$ & $0.001$ & $0$\\
\hline
\end{tabular}
\caption{The absolute values of the hopping integrals between the orbitals of copper atoms at the sites ${\bf{f}} \equiv \left( {{f_x},{f_y}} \right)$ and ${\bf{f}} + {\bf{R}} = \left( {{f_x} + {R_x},{f_y} + {R_y}} \right)$ in the eight-band $p-d$ model Hamiltonian (\ref{eq:Ham_el}).}\label{table4}
\end{table}
\section{The orthogonalization procedure and structural factors}\label{app:str_factors}
To introduce molecular oxygen orbitals orthogonal in adjacent clusters, the Shastry type transformation of in k-space generalized to the case of three oxygen orbitals $p_x$, $p_y$, $p_z$ has the form:
\begin{equation}\label{eq:Shastry}
\left[ {\begin{array}{*{20}{c}}
{{\alpha _{\bf{k}}}}\\
{{\beta _{\bf{k}}}}\\
{{\gamma _{\bf{k}}}}
\end{array}} \right] = \left[ {\begin{array}{*{20}{c}}
{\frac{{{f_{2{\bf{k}}}}{f_{3{\bf{k}}}}}}{{{\mu _{\bf{k}}}{\eta _{\bf{k}}}}}}&{\frac{{{f_{1{\bf{k}}}}{f_{3{\bf{k}}}}}}{{{\mu _{\bf{k}}}{\eta _{\bf{k}}}}}}&{\frac{{{\mu _{\bf{k}}}}}{{{\eta _{\bf{k}}}}}}\\
{\frac{{i{f_{1{\bf{k}}}}}}{{{\mu _{\bf{k}}}}}}&{ - \frac{{i{f_{2{\bf{k}}}}}}{{{\mu _{\bf{k}}}}}}&0\\
{ - \frac{{i{f_{2{\bf{k}}}}}}{{{\eta _{\bf{k}}}}}}&{ - \frac{{i{f_{1{\bf{k}}}}}}{{{\eta _{\bf{k}}}}}}&{\frac{{i{f_{3{\bf{k}}}}}}{{{\eta _{\bf{k}}}}}}
\end{array}} \right]\left[ {\begin{array}{*{20}{c}}
{{p_{x{\bf{k}}}}}\\
{{p_{y{\bf{k}}}}}\\
{{p_{z{\bf{k}}}}}
\end{array}} \right]\,
\end{equation}
where ${\mu _{\bf{k}}} = \sqrt {f_{1{\bf{k}}}^2 + f_{2{\bf{k}}}^2} $, ${\eta _{\bf{k}}} = \sqrt {f_{1{\bf{k}}}^2 + f_{2{\bf{k}}}^2 + f_{3{\bf{k}}}^2} $, and the real functions ${f_{1{\bf{k}}}}$, ${f_{2{\bf{k}}}}$, ${f_{3{\bf{k}}}}$ are chosen as:
\begin{eqnarray}\label{eq:f_functions}
&&{f_{1{\bf{k}}}} = \sin \left( {\frac{{{k_x} + {k_y}}}{2}} \right)\\\nonumber
&&{f_{2{\bf{k}}}} = \sin \left( {\frac{{{k_x} - {k_y}}}{2}} \right)\\\nonumber
&&{f_{3{\bf{k}}}} = \frac{1}{2}\left( {\sin \left( {\frac{{{k_x} + {k_y}}}{2}} \right) + \sin \left( {\frac{{{k_x} - {k_y}}}{2}} \right)} \right)\nonumber
\end{eqnarray}
After introducing molecular orbitals, the Hamiltonian (\ref{eq:Ham_el}) is rewritten in terms of new energy and structure factors. The general definitions of the energies of the oxygen molecular orbitals and the local renormalized hopping integrals in the Hamiltonian (\ref{eq:Ham_mo}) are:
\begin{multline}
\nu _{{\bf{fg}}}^{\left( i \right)} = \frac{1}{N}\sum\limits_{\bf{k}} {\nu _{\bf{k}}^{\left( i \right)}} {e^{ - i{\bf{k}}\left( {{\bf{f}} - {\bf{g}}} \right)}},\,\,i = 1,2,3;\,\,\,\,\,\,\,\kappa _{\bf{R}}^{\left( {\lambda {\rho _j}} \right)} = \frac{1}{N}\sum\limits_{\bf{k}} {\kappa _{\bf{k}}^{\left( {\lambda {\rho _j}} \right)}{e^{i{\bf{kR}}}}} ,\,\,j = 1,2,3,\\
\nu _{{\bf{fg}}}^{\left( 1 \right)} = \frac{{\left( {{\varepsilon _p}f_{3{\bf{k}}}^2 + {\varepsilon _{p_z}}\mu _{\bf{k}}^2} \right)}}{{\eta _{\bf{k}}^2}} + \left( {\frac{{f_{2{\bf{k}}}^2f_{3{\bf{k}}}^2}}{{\eta _{\bf{k}}^2\mu _{\bf{k}}^2}}t_{\bf{k}}^{p_xp_x} + \frac{{f_{1{\bf{k}}}^2f_{3{\bf{k}}}^2}}{{\eta _{\bf{k}}^2\mu _{\bf{k}}^2}}t_{\bf{k}}^{p_yp_y} + \frac{{\mu _{\bf{k}}^2}}{{\eta _{\bf{k}}^2}}t_{\bf{k}}^{p_zp_z}} \right) + \frac{{{f_{1{\bf{k}}}}{f_{2{\bf{k}}}}f_{3{\bf{k}}}^2}}{{\eta _{\bf{k}}^2\mu _{\bf{k}}^2}}\left[ {t_{\bf{k}}^{p_xp_y} + t_{\bf{k}}^{p_yp_x}} \right] + \\
 + \frac{{{f_{2{\bf{k}}}}{f_{3{\bf{k}}}}}}{{\eta _{\bf{k}}^2}}\left[ {t_{\bf{k}}^{p_xp_z} + t_{\bf{k}}^{p_zp_x}} \right] + \frac{{{f_{1{\bf{k}}}}{f_{3{\bf{k}}}}}}{{\eta _{\bf{k}}^2}}\left[ {t_{\bf{k}}^{p_yp_z} + t_{\bf{k}}^{p_zp_y}} \right],\\
\nu _{\bf{k}}^{\left( 2 \right)} = {\varepsilon _p} + \left( {\frac{{f_{1{\bf{k}}}^2}}{{\mu _{\bf{k}}^2}}t_{\bf{k}}^{p_xp_x} + \frac{{f_{2{\bf{k}}}^2}}{{\mu _{\bf{k}}^2}}t_{\bf{k}}^{p_yp_y} - \frac{{{f_{1{\bf{k}}}}{f_{2{\bf{k}}}}}}{{\mu _{\bf{k}}^2}}\left[ {t_{\bf{k}}^{p_xp_y} + t_{\bf{k}}^{p_yp_x}} \right]} \right),\\
\nu _{\bf{k}}^{\left( 3 \right)} = \frac{{\left( {{\varepsilon _p}\mu _{\bf{k}}^2 + {\varepsilon _{p_z}}f_{3{\bf{k}}}^2} \right)}}{{\eta _{\bf{k}}^2}} - \left( {\frac{{f_{2{\bf{k}}}^2}}{{\eta _{\bf{k}}^2}}t_{\bf{k}}^{p_xp_x} + \frac{{f_{1{\bf{k}}}^2}}{{\eta _{\bf{k}}^2}}t_{\bf{k}}^{p_yp_y} + \frac{{f_{3{\bf{k}}}^2}}{{\eta _{\bf{k}}^2}}t_{\bf{k}}^{p_zp_z}} \right) - \frac{{{f_{1{\bf{k}}}}{f_{2{\bf{k}}}}}}{{\eta _{\bf{k}}^2}}\left[ {t_{\bf{k}}^{p_xp_y} + t_{\bf{k}}^{p_yp_x}} \right]\\
 + \frac{{{f_{2{\bf{k}}}}{f_{3{\bf{k}}}}}}{{\eta _{\bf{k}}^2}}\left[ {t_{\bf{k}}^{p_xp_z} + t_{\bf{k}}^{p_zp_x}} \right] + \frac{{{f_{1{\bf{k}}}}{f_{3{\bf{k}}}}}}{{\eta _{\bf{k}}^2}}\left[ {t_{\bf{k}}^{p_yp_z} + t_{\bf{k}}^{p_zp_y}} \right],\\
\nu _{\bf{k}}^{\left( {12} \right)} =  - \frac{{i{f_{1{\bf{k}}}}{f_{2{\bf{k}}}}{f_{3{\bf{k}}}}}}{{{\eta _{\bf{k}}}\mu _{\bf{k}}^2}}t_{\bf{k}}^{p_xp_x} + \frac{{i{f_{1{\bf{k}}}}{f_{2{\bf{k}}}}{f_{3{\bf{k}}}}}}{{{\eta _{\bf{k}}}\mu _{\bf{k}}^2}}t_{\bf{k}}^{p_yp_y} + \frac{{if_{2{\bf{k}}}^2{f_{3{\bf{k}}}}}}{{{\eta _{\bf{k}}}\mu _{\bf{k}}^2}}t_{\bf{k}}^{p_xp_y} - \frac{{if_{1{\bf{k}}}^2{f_{3{\bf{k}}}}}}{{{\eta _{\bf{k}}}\mu _{\bf{k}}^2}}t_{\bf{k}}^{p_yp_x} - \frac{{i{f_{1{\bf{k}}}}}}{{{\eta _{\bf{k}}}}}t_{\bf{k}}^{p_zp_x} + \frac{{i{f_{2{\bf{k}}}}}}{{{\eta _{\bf{k}}}}}t_{\bf{k}}^{p_zp_y},\\
\nu _{\bf{k}}^{\left( {13} \right)} = \frac{{if_{2{\bf{k}}}^2{f_{3{\bf{k}}}}}}{{\eta _{\bf{k}}^2{\mu _{\bf{k}}}}}t_{\bf{k}}^{p_xp_x} + \frac{{if_{1{\bf{k}}}^2{f_{3{\bf{k}}}}}}{{\eta _{\bf{k}}^2{\mu _{\bf{k}}}}}t_{\bf{k}}^{p_yp_y} - \frac{{i{\mu _{\bf{k}}}{f_{3{\bf{k}}}}}}{{\eta _{\bf{k}}^2}}t_{\bf{k}}^{p_zp_z} + \frac{{i{f_{1{\bf{k}}}}{f_{2{\bf{k}}}}{f_{3{\bf{k}}}}}}{{\eta _{\bf{k}}^2{\mu _{\bf{k}}}}}\left[ {t_{\bf{k}}^{p_xp_y} + t_{\bf{k}}^{p_yp_x}} \right] - \frac{{i{f_{2{\bf{k}}}}f_{3{\bf{k}}}^2}}{{\eta _{\bf{k}}^2{\mu _{\bf{k}}}}}t_{\bf{k}}^{p_xp_z} + \\
 + \frac{{i{f_{2{\bf{k}}}}{\mu _{\bf{k}}}}}{{\eta _{\bf{k}}^2}}t_{\bf{k}}^{p_zp_x} + \frac{{i{f_{1{\bf{k}}}}{\mu _{\bf{k}}}}}{{\eta _{\bf{k}}^2}}t_{\bf{k}}^{p_zp_y} - \frac{{i{f_{1{\bf{k}}}}f_{3{\bf{k}}}^2}}{{\eta _{\bf{k}}^2{\mu _{\bf{k}}}}}t_{\bf{k}}^{p_yp_z},\\
\nu _{\bf{k}}^{\left( {23} \right)} =  - \frac{{{f_{1{\bf{k}}}}{f_{2{\bf{k}}}}}}{{{\eta _{\bf{k}}}{\mu _{\bf{k}}}}}t_{\bf{k}}^{p_xp_x} + \frac{{{f_{1{\bf{k}}}}{f_{2{\bf{k}}}}}}{{{\eta _{\bf{k}}}{\mu _{\bf{k}}}}}t_{\bf{k}}^{p_yp_y} + \frac{{f_{2{\bf{k}}}^2}}{{{\eta _{\bf{k}}}{\mu _{\bf{k}}}}}t_{\bf{k}}^{p_yp_x} - \frac{{f_{1{\bf{k}}}^2}}{{{\eta _{\bf{k}}}{\mu _{\bf{k}}}}}t_{\bf{k}}^{p_xp_y} + \frac{{{f_{1{\bf{k}}}}{f_{3{\bf{k}}}}}}{{{\eta _{\bf{k}}}{\mu _{\bf{k}}}}}t_{\bf{k}}^{p_xp_z} - \frac{{{f_{2{\bf{k}}}}{f_{3{\bf{k}}}}}}{{{\eta _{\bf{k}}}{\mu _{\bf{k}}}}}t_{\bf{k}}^{p_yp_z},
\label{eq:energ_mo}
\end{multline}
where $t_{\bf{k}}^{{\lambda}{\lambda '}}$ is the Fourier transform of the hopping integral ${t_{{\lambda}{\lambda '}}}\left( {\bf{R}} \right)$. The local renormalized hopping integrals in the Hamiltonian (\ref{eq:Ham_mo}) are defined as:
\begin{multline}
\kappa _{\bf{k}}^{\left( {d_{{x^2} - {y^2}}\alpha } \right)} = \frac{{{f_{2{\bf{k}}}}{f_{3{\bf{k}}}}}}{{{\mu _{\bf{k}}}{\eta _{\bf{k}}}}}t_{\bf{k}}^{d_{{x^2} - {y^2}}p_x} + \frac{{{f_{1{\bf{k}}}}{f_{3{\bf{k}}}}}}{{{\eta _{\bf{k}}}{\mu _{\bf{k}}}}}t_{\bf{k}}^{d_{{x^2} - {y^2}}p_y} + \frac{{{\mu _{\bf{k}}}}}{{{\eta _{\bf{k}}}}}t_{\bf{k}}^{d_{{x^2} - {y^2}}p_z},\\
\kappa _{\bf{k}}^{\left( {d_{{x^2} - {y^2}}\beta } \right)} =  - \frac{{i{f_{1{\bf{k}}}}}}{{{\mu _{\bf{k}}}}}t_{\bf{k}}^{d_{{x^2} - {y^2}}p_x} + \frac{{i{f_{2{\bf{k}}}}}}{{{\mu _{\bf{k}}}}}t_{\bf{k}}^{d_{{x^2} - {y^2}}p_y},\\
\kappa _{\bf{k}}^{\left( {d_{{x^2} - {y^2}}\gamma } \right)} = \frac{{i{f_{2{\bf{k}}}}}}{{{\eta _{\bf{k}}}}}t_{\bf{k}}^{d_{{x^2} - {y^2}}p_x} + \frac{{i{f_{1{\bf{k}}}}}}{{{\eta _{\bf{k}}}}}t_{\bf{k}}^{d_{{x^2} - {y^2}}p_y} - \frac{{i{f_{3{\bf{k}}}}}}{{{\eta _{\bf{k}}}}}t_{\bf{k}}^{d_{{x^2} - {y^2}}p_z},\,\,\\
\kappa _{\bf{k}}^{\left( {d_{3{z^2} - {r^2}}\alpha } \right)} = \frac{{{f_{2{\bf{k}}}}{f_{3{\bf{k}}}}}}{{{\mu _{\bf{k}}}{\eta _{\bf{k}}}}}t_{\bf{k}}^{d_{3{z^2} - {r^2}}p_x} + \frac{{{f_{1{\bf{k}}}}{f_{3{\bf{k}}}}}}{{{\eta _{\bf{k}}}{\mu _{\bf{k}}}}}t_{\bf{k}}^{d_{3{z^2} - {r^2}}p_y} + \frac{{{\mu _{\bf{k}}}}}{{{\eta _{\bf{k}}}}}t_{\bf{k}}^{d_{3{z^2} - {r^2}}p_z},\\
\kappa _{\bf{k}}^{\left( {d_{3{z^2} - {r^2}}\beta } \right)} =  - \frac{{i{f_{1{\bf{k}}}}}}{{{\mu _{\bf{k}}}}}t_{\bf{k}}^{d_{3{z^2} - {r^2}}p_x} + \frac{{i{f_{2{\bf{k}}}}}}{{{\mu _{\bf{k}}}}}t_{\bf{k}}^{d_{3{z^2} - {r^2}}p_y},\,\\
\kappa _{\bf{k}}^{\left( {d_{3{z^2} - {r^2}}\gamma } \right)} = \frac{{i{f_{2{\bf{k}}}}}}{{{\eta _{\bf{k}}}}}t_{\bf{k}}^{d_{3{z^2} - {r^2}}p_x} + \frac{{i{f_{1{\bf{k}}}}}}{{{\eta _{\bf{k}}}}}t_{\bf{k}}^{d_{3{z^2} - {r^2}}p_y} - \frac{{i{f_{3{\bf{k}}}}}}{{{\eta _{\bf{k}}}}}t_{\bf{k}}^{d_{3{z^2} - {r^2}}p_z}\\
\kappa _{\bf{k}}^{\left( {d_{xy}\alpha } \right)} = \frac{{{f_{2{\bf{k}}}}{f_{3{\bf{k}}}}}}{{{\mu _{\bf{k}}}{\eta _{\bf{k}}}}}t_{\bf{k}}^{d_{xy}p_x} + \frac{{{f_{1{\bf{k}}}}{f_{3{\bf{k}}}}}}{{{\eta _{\bf{k}}}{\mu _{\bf{k}}}}}t_{\bf{k}}^{d_{xy}p_y} + \frac{{{\mu _{\bf{k}}}}}{{{\eta _{\bf{k}}}}}t_{\bf{k}}^{d_{xy}p_z},\,\kappa _{\bf{k}}^{\left( {d_{xy}\beta } \right)} =  - \frac{{i{f_{1{\bf{k}}}}}}{{{\mu _{\bf{k}}}}}t_{\bf{k}}^{d_{xy}p_x} + \frac{{i{f_{2{\bf{k}}}}}}{{{\mu _{\bf{k}}}}}t_{\bf{k}}^{d_{xy}p_y},\\
\kappa _{\bf{k}}^{\left( {d_{xy}\gamma } \right)} = \frac{{i{f_{2{\bf{k}}}}}}{{{\eta _{\bf{k}}}}}t_{\bf{k}}^{d_{xy}p_x} + \frac{{i{f_{1{\bf{k}}}}}}{{{\eta _{\bf{k}}}}}t_{\bf{k}}^{d_{xy}p_y} - \frac{{i{f_{3{\bf{k}}}}}}{{{\eta _{\bf{k}}}}}t_{\bf{k}}^{d_{xy}p_z},\,\\
\kappa _{\bf{k}}^{\left( {d_{xz}\alpha } \right)} = \frac{{{f_{2{\bf{k}}}}{f_{3{\bf{k}}}}}}{{{\mu _{\bf{k}}}{\eta _{\bf{k}}}}}t_{\bf{k}}^{d_{xz}p_x} + \frac{{{f_{1{\bf{k}}}}{f_{3{\bf{k}}}}}}{{{\eta _{\bf{k}}}{\mu _{\bf{k}}}}}t_{\bf{k}}^{d_{xz}p_y} + \frac{{{\mu _{\bf{k}}}}}{{{\eta _{\bf{k}}}}}t_{\bf{k}}^{d_{xz}p_z},\,\kappa _{\bf{k}}^{\left( {d_{xz}\beta } \right)} =  - \frac{{i{f_{1{\bf{k}}}}}}{{{\mu _{\bf{k}}}}}t_{\bf{k}}^{d_{xz}p_x} + \frac{{i{f_{2{\bf{k}}}}}}{{{\mu _{\bf{k}}}}}t_{\bf{k}}^{d_{xz}p_y},\,\\
\kappa _{\bf{k}}^{\left( {d_{xz}\gamma } \right)} = \frac{{i{f_{2{\bf{k}}}}}}{{{\eta _{\bf{k}}}}}t_{\bf{k}}^{d_{xz}p_x} + \frac{{i{f_{1{\bf{k}}}}}}{{{\eta _{\bf{k}}}}}t_{\bf{k}}^{d_{xz}p_y} - \frac{{i{f_{3{\bf{k}}}}}}{{{\eta _{\bf{k}}}}}t_{\bf{k}}^{d_{xz}p_z},\,\,\,\\
\kappa _{\bf{k}}^{\left( {d_{yz}\alpha } \right)} = \frac{{{f_{2{\bf{k}}}}{f_{3{\bf{k}}}}}}{{{\mu _{\bf{k}}}{\eta _{\bf{k}}}}}t_{\bf{k}}^{d_{yz}p_x} + \frac{{{f_{1{\bf{k}}}}{f_{3{\bf{k}}}}}}{{{\eta _{\bf{k}}}{\mu _{\bf{k}}}}}t_{\bf{k}}^{d_{yz}p_y} + \frac{{{\mu _{\bf{k}}}}}{{{\eta _{\bf{k}}}}}t_{\bf{k}}^{d_{yz}p_z},\kappa _{\bf{k}}^{\left( {d_{yz}\beta } \right)} =  - \frac{{i{f_{1{\bf{k}}}}}}{{{\mu _{\bf{k}}}}}t_{\bf{k}}^{d_{yz}p_x} + \frac{{i{f_{2{\bf{k}}}}}}{{{\mu _{\bf{k}}}}}t_{\bf{k}}^{d_{yz}p_y},\\
\kappa _{\bf{k}}^{\left( {d_{yz}\gamma } \right)} = \frac{{i{f_{2{\bf{k}}}}}}{{{\eta _{\bf{k}}}}}t_{\bf{k}}^{d_{yz}p_x} + \frac{{i{f_{1{\bf{k}}}}}}{{{\eta _{\bf{k}}}}}t_{\bf{k}}^{d_{yz}p_y} - \frac{{i{f_{3{\bf{k}}}}}}{{{\eta _{\bf{k}}}}}t_{\bf{k}}^{d_{yz}p_z},\,\,\,
\label{eq:ren_hop}
\end{multline}
The general definitions of the structural factors $\Psi _{{\bf{ff'gg'}}}^{{\rho _i}{\rho _{i'}}{\rho _j}{\rho _{j'}}} $ and $\Phi _{{\bf{fgg'}}}^{{\rho _i}{\rho _j}}$ in the Hamiltonian (\ref{eq:Ham_mo}) are:
\begin{multline}
\Psi _{{\bf{ff'gg'}}}^{{\rho _i}{\rho _{i'}}{\rho _j}{\rho _{j'}}} =  \frac{1}{{{N^3}}}\sum\limits_{\zeta \zeta '} {\sum\limits_{{\bf{kqm}}} {S_{i\zeta {\bf{k}}}^*{S_{i'\zeta {\bf{q}}}}S_{j\zeta '{\bf{m}}}^*{S_{j'\zeta '\left( {{\bf{k}} - {\bf{q}} + {\bf{m}}} \right)}}{e^{ - i{\bf{k}}\left( {{\bf{f}} - {\bf{g'}}} \right)}}{e^{i{\bf{q}}\left( {{\bf{f'}} - {\bf{g'}}} \right)}}{e^{ - i{\bf{m}}\left( {{\bf{g}} - {\bf{g'}}} \right)}}} } \\
{S_{1{p_x}{\bf{k}}}} \equiv {S_{\alpha {p_x}{\bf{k}}}} = \frac{{{f_{2{\bf{k}}}}{f_{3{\bf{k}}}}}}{{{\eta _{\bf{k}}}{\mu _{\bf{k}}}}},\,\,\,{S_{2{p_x}{\bf{k}}}} \equiv {S_{\beta {p_x}{\bf{k}}}} =  - \frac{{i{f_{1{\bf{k}}}}}}{{{\mu _{\bf{k}}}}},\,\,\,{S_{3{p_x}{\bf{k}}}} \equiv {S_{\gamma {p_x}{\bf{k}}}} = \frac{{i{f_{2{\bf{k}}}}}}{{{\eta _{\bf{k}}}}}\\
{S_{1{p_y}{\bf{k}}}} = \frac{{{f_{1{\bf{k}}}}{f_{3{\bf{k}}}}}}{{{\eta _{\bf{k}}}{\mu _{\bf{k}}}}},\,\,{S_{2{p_y}{\bf{k}}}} = \frac{{i{f_{2{\bf{k}}}}}}{{{\mu _{\bf{k}}}}},\,\,\,{S_{3{p_y}{\bf{k}}}} = \frac{{i{f_{1{\bf{k}}}}}}{{{\eta _{\bf{k}}}}}\,\\
{S_{1{p_z}{\bf{k}}}} = \frac{{{\mu _{\bf{k}}}}}{{{\eta _{\bf{k}}}}},\,\,\,\,\,\,{S_{2{p_z}{\bf{k}}}} = 0,\,\,\,\,\,\,{S_{3{p_z}{\bf{k}}}} =  - \frac{{i{f_{3{\bf{k}}}}}}{{{\eta _{\bf{k}}}}}\,\,\\
\Phi _{{\bf{fgg'}}}^{{\rho _i}{\rho _j}} = \frac{1}{{{N^2}}}\sum\limits_{{\bf{kq}}} {\Phi _{{\bf{kq}}}^{{\rho _i}{\rho _j}}{C_{{\bf{kq}}}}\left( {{\bf{f}},{\bf{g}},{\bf{g'}}} \right)} ,\\
{C_{{\bf{kq}}}}\left( {{\bf{f}},{\bf{g}},{\bf{g'}}} \right) = 2\left[ {\cos \left( {\left( {{\bf{k}} - {\bf{q}}} \right)\left( {{a \mathord{\left/
 {\vphantom {a {2 + {b \mathord{\left/
 {\vphantom {b 2}} \right.
 \kern-\nulldelimiterspace} 2}}}} \right.
 \kern-\nulldelimiterspace} {2 + {b \mathord{\left/
 {\vphantom {b 2}} \right.
 \kern-\nulldelimiterspace} 2}}}} \right)} \right) + \cos \left( {\left( {{\bf{k}} - {\bf{q}}} \right)\left( {{a \mathord{\left/
 {\vphantom {a {2 - {b \mathord{\left/
 {\vphantom {b 2}} \right.
 \kern-\nulldelimiterspace} 2}}}} \right.
 \kern-\nulldelimiterspace} {2 - {b \mathord{\left/
 {\vphantom {b 2}} \right.
 \kern-\nulldelimiterspace} 2}}}} \right)} \right)} \right]{e^{ - i{\bf{k}}\left( {{\bf{g}} - {\bf{f}}} \right)}}{e^{i{\bf{q}}\left( {{\bf{g'}} - {\bf{f}}} \right)}},\\
\Phi _{{\bf{kq}}}^{\alpha \alpha } = \frac{{{f_{2{\bf{k}}}}{f_{3{\bf{k}}}}}}{{{\eta _{\bf{k}}}{\mu _{\bf{k}}}}}\frac{{{f_{2{\bf{n}}}}{f_{3{\bf{n}}}}}}{{{\eta _{\bf{n}}}{\mu _{\bf{n}}}}} + \frac{{{f_{1{\bf{k}}}}{f_{3{\bf{k}}}}}}{{{\eta _{\bf{k}}}{\mu _{\bf{k}}}}}\frac{{{f_{1{\bf{n}}}}{f_{3{\bf{n}}}}}}{{{\eta _{\bf{n}}}{\mu _{\bf{n}}}}} + \frac{{{\mu _{\bf{k}}}}}{{{\eta _{\bf{k}}}}}\frac{{{\mu _{\bf{n}}}}}{{{\eta _{\bf{n}}}}},\,\,\,\Phi _{{\bf{kq}}}^{\beta \beta } = \left[ {\frac{{{f_{1{\bf{k}}}}}}{{{\mu _{\bf{k}}}}}\frac{{{f_{1{\bf{q}}}}}}{{{\mu _{\bf{q}}}}} + \frac{{{f_{2{\bf{k}}}}}}{{{\mu _{\bf{k}}}}}\frac{{{f_{2{\bf{q}}}}}}{{{\mu _{\bf{q}}}}}} \right],\\
\Phi _{{\bf{kq}}}^{\gamma \gamma } = \left[ {\frac{{{f_{2{\bf{k}}}}}}{{{\eta _{\bf{k}}}}}\frac{{{f_{2{\bf{q}}}}}}{{{\eta _{\bf{q}}}}} + \frac{{{f_{1{\bf{k}}}}}}{{{\eta _{\bf{k}}}}}\frac{{{f_{1{\bf{q}}}}}}{{{\eta _{\bf{q}}}}} + \frac{{{f_{3{\bf{k}}}}}}{{{\eta _{\bf{k}}}}}\frac{{{f_{3{\bf{q}}}}}}{{{\eta _{\bf{q}}}}}} \right],\\
\Phi _{{\bf{kq}}}^{\alpha \beta } = \left[ { - \frac{{{f_{2{\bf{k}}}}{f_{3{\bf{k}}}}}}{{{\eta _{\bf{k}}}{\mu _{\bf{k}}}}}\frac{{i{f_{1{\bf{q}}}}}}{{{\mu _{\bf{q}}}}} + \frac{{{f_{1{\bf{k}}}}{f_{3{\bf{k}}}}}}{{{\eta _{\bf{k}}}{\mu _{\bf{k}}}}}\frac{{i{f_{2{\bf{q}}}}}}{{{\mu _{\bf{q}}}}}} \right],\Phi _{{\bf{kq}}}^{\beta \alpha } = \left[ {\frac{{i{f_{1{\bf{k}}}}}}{{{\mu _{\bf{k}}}}}\frac{{{f_{2{\bf{q}}}}{f_{3{\bf{q}}}}}}{{{\eta _{\bf{q}}}{\mu _{\bf{q}}}}} - \frac{{i{f_{2{\bf{k}}}}}}{{{\mu _{\bf{k}}}}}\frac{{{f_{1{\bf{q}}}}{f_{3{\bf{q}}}}}}{{{\eta _{\bf{q}}}{\mu _{\bf{q}}}}}} \right],\\
\Phi _{{\bf{kq}}}^{\alpha \gamma } = \left[ {\frac{{{f_{2{\bf{k}}}}{f_{3{\bf{k}}}}}}{{{\eta _{\bf{k}}}{\mu _{\bf{k}}}}}\frac{{i{f_{2{\bf{q}}}}}}{{{\eta _{\bf{q}}}}} + \frac{{{f_{1{\bf{k}}}}{f_{3{\bf{k}}}}}}{{{\eta _{\bf{k}}}{\mu _{\bf{k}}}}}\frac{{i{f_{1{\bf{q}}}}}}{{{\eta _{\bf{q}}}}} - \frac{{{\mu _{\bf{k}}}}}{{{\eta _{\bf{k}}}}}\frac{{i{f_{3{\bf{q}}}}}}{{{\eta _{\bf{q}}}}}} \right],\,\,\Phi _{{\bf{kq}}}^{\gamma \alpha } = \left[ { - \frac{{i{f_{2{\bf{k}}}}}}{{{\eta _{\bf{k}}}}}\frac{{{f_{2{\bf{q}}}}{f_{3{\bf{q}}}}}}{{{\eta _{\bf{q}}}{\mu _{\bf{q}}}}} - \frac{{i{f_{1{\bf{k}}}}}}{{{\eta _{\bf{k}}}}}\frac{{{f_{1{\bf{q}}}}{f_{3{\bf{q}}}}}}{{{\eta _{\bf{q}}}{\mu _{\bf{q}}}}} + \frac{{i{f_{3{\bf{k}}}}}}{{{\eta _{\bf{k}}}}}\frac{{{\mu _{\bf{q}}}}}{{{\eta _{\bf{q}}}}}} \right],\\
\Phi _{{\bf{kq}}}^{\beta \gamma } = \left[ { - \frac{{{f_{1{\bf{k}}}}}}{{{\mu _{\bf{k}}}}}\frac{{{f_{2{\bf{q}}}}}}{{{\eta _{\bf{q}}}}} + \frac{{{f_{2{\bf{k}}}}}}{{{\mu _{\bf{k}}}}}\frac{{{f_{1{\bf{q}}}}}}{{{\eta _{\bf{q}}}}}} \right],\,\,\Phi _{{\bf{kq}}}^{\gamma \beta } = \left[ { - \frac{{{f_{2{\bf{k}}}}}}{{{\eta _{\bf{k}}}}}\frac{{{f_{1{\bf{q}}}}}}{{{\mu _{\bf{q}}}}} + \frac{{{f_{1{\bf{k}}}}}}{{{\eta _{\bf{k}}}}}\frac{{{f_{2{\bf{q}}}}}}{{{\mu _{\bf{q}}}}}} \right].
\label{eq:HUp_Psi}
\end{multline}
%%       \citet{<label>} ==> Lamport (1994)
%%       \citep{<label>} ==> (Lamport, 1994)
%%
%Example citation, See \citet{lamport94}.




%

\end{document}